\documentclass[pra,twocolumn,showpacs,superscriptaddress,floatfix,nofootinbib]{revtex4-1}
\usepackage{epsfig}
\usepackage{latexsym}
\usepackage{amsmath}
\usepackage{graphics}
\usepackage{float}

\usepackage{color}
\usepackage{pstricks}

\newcommand{\br}{\mathbf{r}}

\newcommand{\bq}{\mathbf{q}}
\newcommand{\brp}{\mathbf{r}_\parallel}

\newcommand{\bqp}{\mathbf{q}_\parallel}
\newcommand{\TE}{{\rm TE}}
\newcommand{\TM}{{\rm TM}}
\newcommand{\T}{\textsf{T}}
\newcommand{\w}{\omega}
\newcommand{\rd}{\hspace{-1 mm}{ d}}
\begin{document}

\title{Quantum electrodynamics near a Huttner-Barnett dielectric}
\author{Claudia Eberlein}
\author{Robert Zietal}
\affiliation{Department of Physics \& Astronomy,
    University of Sussex,
     Falmer, Brighton BN1 9QH, England}
\date{\today}
\begin{abstract}
We build up a consistent theory of quantum electrodynamics in the presence of macroscopic polarizable media. We use the Huttner-Barnett model of a dispersive and absorbing dielectric medium and formulate the theory in terms of interacting quantum fields. We integrate out the damped polaritons by using diagrammatic techniques and find an exact expression for the displacement field (photon) propagator in the presence of a dispersive and absorbing dielectric half-space. This opens a new route to traceable perturbative calculations of the same kind as in free-space quantum electrodynamics. As a worked-through example we consider the interaction of a neutral atom with a dispersive and absorbing dielectric half-space. For that we use the multipolar coupling $\boldsymbol{\mu}\cdot\mathbf{D}$ of the atomic dipole moment to the electromagnetic displacement field. 
We apply the newly developed formalism to calculate the one-loop correction to the atomic electron propagator and find the energy-level shift and changes in the spontaneous decay rates for a neutral atom close to an absorptive dielectric mirror. 
\end{abstract}

\pacs{31.70.-f, 41.20.Cv, 42.50.Pq}

\maketitle
\section{\label{sec:level1}Introduction.}
Quantum electrodynamics is a well-functioning theory which accurately predicts a wide range of phenomena, not just in high-energy physics but also in atomic physics. The best known quantum electrodynamic effect in atomic physics is certainly the Lamb shift which has by now been calculated to very high accuracy \cite{Pachucki}. If the atom is located not in free space but instead near a reflecting surface, which could be dielectric or conducting, then the reflection of photons from that surface leads to the Lamb shift acquiring a distance-dependent component, the Casimir-Polder shift, whose gradient yields the Casimir-Polder force between atom and surface.
Alternatively, the Casimir-Polder shift can be viewed as a Stark effect where the role of the electric field is played by the non-zero and position-dependent electromagnetic vacuum fluctuations in the presence of dielectrics \cite{Milonni}. 
In order to study the Casimir-Polder effect and related quantum electrodynamic effects due to the presence of macroscopic material boundaries, one needs a theory of the quantized electromagnetic field in the presence of such boundaries. The method of field quantization largely depends on how sophisticated a model of the material's optical response one assumes. In the simplest case one might assume perfect reflectivity of the surface. 
The quantization of the electromagnetic field can then be achieved quite easily by a normal-mode expansion of the field, where the electromagnetic field is expanded in terms of a complete set of solutions of the homogeneous Helmholtz equation. The presence of the boundaries is taken into account by imposing appropriate boundary conditions on the electromagnetic field. Quantization is then accomplished by promoting the expansion coefficients of each mode to creation and annihilation operators which are required to satisfy bosonic commutation relations. This approach of canonical quantization has the advantage of being simple and therefore workable even for complex geometries \cite{Cylinder} 
but the perfect-reflector model for the surface lacks essential physical features, e.g. evanescent modes, which may have a dramatic effect on predicted quantities \cite{Electron}.
An improvement is to consider the material as a non-dispersive and non-absorbing dielectric characterized by a single real number, an index of refraction.
Then field quantization can still be achieved by canonical quantization using field modes, although the specific implementation of the method requires a lot more care than for perfect reflectors \cite{Glauber}. 

Canonical quantization of the electromagnetic field in terms of normal modes runs into difficulties when one wants to include in the formalism realistic properties of dielectrics. The response of the material's surface to the electromagnetic radiation in reality depends on the frequency of the impinging radiation. Furthermore, causality requirements demand that any dispersion is always accompanied by absorption. However, a naive incorporation of absorption into canonical field quantization leads to field commutators decaying in time, i.e.~an inconsistent theory. Therefore, in any model of interaction between real dielectrics and the electromagnetic field, the field has to be coupled to a reservoir in order to simulate the absorptive degrees of freedom \cite{Huttner1,Huttner11}. This can be done in a number of ways. One is to model the absorptive degrees of freedom by adding to the operator-valued Maxwell equations Langevin-type fluctuating noise-currents that ensure that the canonical commutation relations do not decay in time but rather take the expected form \cite{PhenQED}. In this approach the field equations are solved by using the Green's function of the wave equation, and the noise-current operators and their properties play a major role in describing the dynamics of the coupled field-dielectric system. A number of papers have provided an \textit{a posteriori} microscopic justification of such a procedure by deriving the commutative properties of the noise-current operators that were otherwise introduced \textit{ad hoc} \cite{SuttorpWubs,SuttorpWonderen,Omar}.

A more direct approach to modelling the interaction between the electromagnetic field and an absorptive dielectric is to explicitly include from the outset in the Lagrangian (or Hamiltonian) the matter degrees of freedom that are responsible for absorption. The dielectric is then envisaged as consisting of a continuum of harmonic oscillators coupled to a reservoir which consists of yet another set of harmonic oscillators. This quantum model of a classical dielectric was originally introduced by Hopfield \cite{Hopfield}. The first Fano-type diagonalization \cite{Fano} of the resulting Hamiltonian was achieved for fields in three dimensions in Ref.~\cite{Huttner} for a bulk dielectric and the general treatment of inhomogeneous dielectrics followed in Ref.~\cite{SuttorpWonderen}. This model has also been extended to include spatial dispersion \cite{AdamBechler} and magnetodielectrics \cite{HBMagnet}. Practical applications of the Huttner-Barnett model, e.g. the calculation of spontaneous decay rates \cite{HBRates}, work well for bulk dielectrics where simple forms of the relevant operators can be found, though an additional difficulty is that in a bulk medium local field corrections play an important role and need to be included. On the other hand, complications that arise due to inhomogeneities of the dielectric have previously led to unwieldy and impractical results; the conceptually very interesting work by Yeung and Gustafson \cite{Yeung} uses the Wiener-Hopf method to calculate the photon propagator of the vector field ${\bf A}$ in the presence of an absorbing dielectric half-space, but the result is so complicated that it has to be Fourier transformed and evaluated numerically, whence all subsequent calculations are also necessarily only numerical.

In this paper we demonstrate that by starting from a Power-Zienau-Wooley type of Hamiltonian rather than adopting minimal coupling, one can carry out explicit and easy-to-follow perturbative calculations in quantum electrodynamics in the presence of an inhomogeneous Huttner-Barnett dielectrics. We apply the formalism we develop to the problem of calculating the energy-level shifts and change in spontaneous-decay rates for a neutral atom placed in the vicinity of a dielectric half-space. We successfully re-derive the well-known results of phenomenological methods and broaden them by providing the asymptotic expansions that quantify the influence of absorption on the standard Casimir-Polder force calculated in Ref.~\cite{Wu}. We use only standard methods of quantum field theory, in a similar way as this is done in condensed matter theories. 
This requires the calculation of quantum propagators, most notably that of the electromagnetic field. We show that this task is non-trivial but manageable. Inspired by the results of Ref.~\cite{Valeri} we find an exact solution of the Dyson equation satisfied by the photon propagator. In Appendix~\ref{App:DressPropPhen} we make contact with the phenomenological noise-current approach and calculate the photon propagator using the electromagnetic field operators constructed on the basis of the noise-current operators \cite{PhenQED}.

\section{\label{sec:level2}Construction of the model and Hamiltonians}
We are aiming to study the electromagnetic interaction between a quantum system, e.g. an atom, and a macroscopic absorbing dielectric body. To this end we use the model of absorbing dielectrics developed in \cite{Huttner1} but generalized to inhomogeneous dielectrics. The dielectric is modelled by a continuum of quantized harmonic oscillators -- the polarization field. This, in turn, is coupled to another set of quantized harmonic oscillators -- the reservoir, the presence of which leads to damping in the polarization field so as to allow the absorption of radiation. These coupled quantum fields interact with the electromagnetic field via the coupling of the polarization field to the electric field. It turns out that the subsystem consisting of the reservoir, the polarization and the electromagnetic field is exactly soluble, at least for simple geometries of the dielectric. Therefore, the interaction of the atom with the dielectric can de facto be reduced to the interaction of the atomic dipole with the 'dressed' electromagnetic field, that is the electromagnetic field corrected for the presence of an absorptive body. This approach builds on the theory developed in \cite{SmallBody} where the interaction between an atom and a point-like absorptive dielectric (i.e. damped harmonic oscillator) was addressed. The crucial difference is that for description of the interaction with a point-like absorber the 'dressed' electromagnetic field is required only perturbatively, but in the case of an extended absorbing body one needs to find the 'dressed' electromagnetic field exactly if one wants to capture accurately the interaction with an atom or other quantum system.

Our starting point is the Lagrangian density describing the complete dynamics of the electromagnetic field and the dielectric,
\begin{equation}
\mathcal{L}_0=\mathcal{L}_{\rm EM}+\mathcal{L}_{\rm P}+\mathcal{L}_{\rm R}+\mathcal{L}_{\rm P-EM}\;.
\end{equation}
The various constituent parts are:

(i) \emph{The Lagrangian density $\mathcal{L}_{\rm EM}$ of the free electromagnetic field}:
	\begin{equation}
	\mathcal{L}_{\rm EM}=\frac{\epsilon_0}{2}\mathbf{E}^2(\br)-\frac{1}{2\mu_0}\mathbf{B}^2(\br),
	\label{EMField}
	\end{equation}
	where $\mathbf{E}(\br)$ is the electric field and $\mathbf{B}(\br)$ is the magnetic induction \cite{fn1}. 

(ii) \emph{The Lagrangian density $\mathcal{L}_{\rm P}$ of the polarization field}:
	\begin{eqnarray}
	\mathcal{L}_{\rm P}= \frac{1}{2}\mathcal{M}\dot{\mathbf{X}}(\br)-\frac{1}{2}\mathcal{M}\w_{\rm T}^2\mathbf{X}^2(\br).
	\label{eqn:Polarization}
	\end{eqnarray}
	The field $\mathbf{X}$ is the dipole moment density of the continuum of harmonic oscillators describing the dielectric. The strength of the restoring force acting on the polarization oscillators is determined by the combination $\mathcal{M}\w_{\rm T}^2$. Hence, for a fixed absorption frequency $\w_{\rm T}$ of the dielectric, the 'mass' $\mathcal{M}$ is the parameter that determines the susceptibility of the polarization oscillator to an external agent. It has dimensions of $({\rm mass})\times({\rm length})^{-1}\times({\rm dipole\; moment\; density})^{-2}$. In fact, the quantity $(\mathcal{M}\epsilon_0\omega^2_{\rm T})^{-1}$ will turn out to be the polarizability of the dielectric at zero frequency \cite{Hopfield}. The absence of derivatives with respect to $\br$ in Eq. (\ref{eqn:Polarization}) implies that the polarization oscillators at different points in space are mutually independent resulting in a model with no spatial dispersion. 

(iii) \emph{The Lagrangian density $\mathcal{L}_{\rm R}$ of the reservoir}, including its coupling to the polarization field:
	\begin{eqnarray}
	\mathcal{L}_{\rm R}=\int_0^\infty\rd
	\nu\left\{\frac{1}{2}\rho_\nu\dot{\mathbf{Y}}_\nu(\br)-\frac{1}{2}\rho_\nu\nu^2\left[\mathbf{Y}_\nu(\br)-\mathbf{X}(\br)\right]^2\right\}.\nonumber\\
	\label{eqn:Reservoir}
	\end{eqnarray}
	The set of fields $\mathbf{Y}_\nu$ represent the dipole moment density of the bath oscillators at all bath frequencies $\nu$, and the parameter $\rho_\nu$ has dimensions of $({\rm mass})\times({\rm length})^{-1}\times({\rm dipole\; moment\; density})^{-2}\times({\rm frequency})^{-1}$. The coupling of the bath to the polarization field leads to the appearance of term proportional to $\dot{\mathbf{X}}(\br,t)$ in the equations of motion for the polarization field; hence it is responsible for damping \cite{Senitzky,Langevin} (cf. also Appendix \ref{App:Epsilon}). The 'masses' of the bath oscillators $\rho_\nu$ vary continuously with index $\nu$ and describe the strength of the coupling between a single polarization oscillator and the continuum of reservoir oscillators for different frequencies $\nu$. The precise profile of $\rho_\nu$ is chosen so that the desired absorption spectrum is obtained \cite{Acta}.
			
(iv) \emph{The Lagrangian density $\mathcal{L}_{\rm P-EM}$ describing the interaction of the polarization field with the electromagnetic field}:
	\begin{equation}
	\mathcal{L}_{\rm P-EM}=-g(\br)\mathbf{X}(\br)\cdot\mathbf{E}(\br).
	\label{eqn:PolField}
	\end{equation}

	The dimensionless coupling function $g(\br)$ specifies where the interaction is taking place, i.e.
	\begin{eqnarray}
	g(\br)=
	\left\{
	\begin{array}{ll}
	1 & {\rm inside\;the\;dielectric}\\
	0 &  {\rm outside\;the\;dielectric}
	\end{array}\right.\;.
	\label{eqn:Gfun}
	\end{eqnarray}
	Thus $g(\br)$ describes the geometric shape of the dielectric body and limits the interaction to its interior. Therefore it is inconsequential whether the polarization field $\mathbf{X}(\br)$ is defined in the whole of space or only in the interior, but the latter would cause unnecessary technical complications later on.

It is straightforward to identify the canonical momenta and obtain the corresponding Hamiltonian densities
\begin{eqnarray}
 \mathcal{H}_{\rm EM}\;&=&\;\frac{1}{2\epsilon_0}\mathbf{D}^2(\br)+\frac{1}{2\mu_0}\mathbf{B}^2(\br),\label{eqn:EMHam}\\
\mathcal{H}_{\rm P}\;&=&\;\frac{\mathbf{P}^2(\br)}{2\mathcal{M}}+\frac{1}{2}\mathcal{M}\w_{\rm T}^2\mathbf{X}^2(\br),\label{eqn:PolarizationHam}\\
\mathcal{H}_{\rm R}\;&=&\;\int_0^\infty\rd \nu\left[\frac{\mathbf{Z}^2_\nu(\br)}{2\rho_\nu}+\frac{1}{2}\rho_\nu\nu^2\mathbf{Y}_\nu^2(\br)\right],\label{eqn:BathHam}\\
\mathcal{H}_{\rm P-R}\;&=&\;-\int_0^\infty\rd \nu\rho_\nu \nu^2\mathbf{X}(\br)\cdot\mathbf{Y}_\nu(\br),\label{eqn:BathResCoupling}\\
\mathcal{H}_{\rm P-EM}\;&=&\;-\frac{g(\br)}{\epsilon_0}\mathbf{D}(\br)\cdot\mathbf{X}(\br).
\label{eqn:PolEMCoupling}
\end{eqnarray}
For convenience we have separated out the polarization-field reservoir coupling $\mathcal{H}_{\rm P-R}$ and also the part of the Hamiltonian that just shifts the eigenfrequency of the polarization field, 
\begin{equation}
\mathcal{H}_{\rm S}=\frac{1}{2}\int_0^\infty\rd \nu\rho_\nu\nu^2\mathbf{X}^2(\br)+\frac{1}{2}\frac{g^2(\br)}{\epsilon_0}\mathbf{X}^2(\br).\label{eqn:FrequencyShifts}
\end{equation}
The first term of (\ref{eqn:FrequencyShifts}) arises due the coupling between the polarization field and the reservoir, whereas the second term is caused by the coupling between the electromagnetic and polarization fields. Equations (\ref{eqn:EMHam})--(\ref{eqn:FrequencyShifts}), accompanied by the set of the equal-time commutation relations
\begin{eqnarray}
 \left[D_i(\br),B_j(\br')\right]\;\;&=\;\;&i\hbar\epsilon^{ijm}\nabla'_m\delta^{(3)}(\br-\br'),\label{eqn:EMCommutator}\\
 \left[X_i(\br),P_j(\br')\right]\;\;&=\;\;&i\hbar\delta_{ij}\delta^{(3)}(\br-\br'),\label{eqn:PolComm}\\
\left[Y_{i,\nu}(\br),Z_{j,\nu'}(\br')\right]\;\;&=\;\;&i\hbar\delta_{ij}\delta^{(3)}(\br-\br')\delta(\nu-\nu'),\label{eqn:BathCommutator}
\end{eqnarray}
allow one to derive the equations of motion for the inhomogeneous damped-polariton model, cf. \cite{SuttorpWubs}. The dielectric displacement ${\mathbf{D}(\br)\equiv\epsilon_0\mathbf{E}(\br)+g(\br)\mathbf{X}(\br)}$ is the negative of the momentum conjugate to electromagnetic vector potential $\mathbf{A}(\br)$, 
as it should be \cite{CraigThiru}. This is assured by the correct choice of coupling (\ref{eqn:PolField}). As already mentioned, the Hamiltonian density $\mathcal{H}_{\rm S}$, Eq. (\ref{eqn:FrequencyShifts}), shifts the eigenfrequency $\w_{\rm T}$ of the polarization field, i.e.
\begin{equation}
 \w_{\rm T}^2\longrightarrow\tilde{\w}_{\rm T}^2=\w_{\rm T}^2+\frac{1}{\mathcal{M}}\int_0^\infty\rd \nu\rho_\nu\nu^2+\frac{g^2(\br)}{\epsilon_0\mathcal{M}}.\label{eqn:FrequencyShift1}
\end{equation}
The second term contains the parameter $\rho_\nu$ that pertains to the shape of the absorption spectrum. For our choice of $\nu$-dependence (see Appendix \ref{App:Epsilon}), it turns out to be infinite. However, this is not problematic as the equations of motion for the fields and hence all observable quantities, most notably the dielectric function, stay finite and physically meaningful. Furthermore, the last term of Eq. (\ref{eqn:FrequencyShift1}) in principle introduces a position-dependence of the frequency $\tilde{\w}_{\rm T}$ through the coupling function $g(\br)$. While not yet apparent at this stage, this position-dependence will turn out irrelevant. Hence, for now we set $g(\br)=1$ in the expression for the frequency shift but shall explain later on why we are allowed to do so. With that we can incorporate equation (\ref{eqn:FrequencyShifts}) into the Hamiltonian density of the polarization field and write
\begin{eqnarray}
 \mathcal{H}_{\rm P}&=&\frac{\mathbf{P}^2(\br)}{2\mathcal{M}}+\frac{1}{2}\mathcal{M}\tilde{\w}_{\rm T}^2\mathbf{X}^2(\br),\label{eqn:PolarizationHam2}
\end{eqnarray}
with
\begin{equation}
\tilde{\w}_{\rm T}^2=\w_{\rm T}^2+\w_{\rm P}^2+\frac{1}{\mathcal{M}}\int_0^\infty\rd \nu\rho_\nu\nu^2 ,\label{eqn:FrequencyShift2}
\end{equation}
where by hindsight, we have introduced the symbol $\w_{\rm P}^2=(\epsilon_0\mathcal{M})^{-1}$ in analogy to the plasma frequency in metals \cite{fn2}.

Our aim  is to investigate the influence of an absorbing dielectric on the properties of an atom, such as the shifts in its energy levels and spontaneous decay rates. 
The complete Hamiltonian, including the atom, reads 
\begin{eqnarray}
H=\int\rd^3\br\left(\mathcal{H}_{\rm A}+
\mathcal{H}_{\rm EM}+\mathcal{H}_{\rm P}+\mathcal{H}_{\rm R}\right.\hspace{2 cm}\nonumber\\
\left.
+\mathcal{H}_{\rm P-R}+\mathcal{H}_{\rm P-EM}+\mathcal{H}_{\rm A-EM}
\right),\hspace{.5 cm}\label{eqn:TotalH}
\end{eqnarray}
where $\mathcal{H}_{A}$ is the Hamiltonian density of the atom and $\mathcal{H}_{A-EM}$ describes its coupling to the electromagnetic field. We consider a one-electron atom and treat the atomic electron non-relativistically by representing it as a quantum of the Schr\"odinger field satisfying fermionic anticommutation relations. The Hamiltonian density $\mathcal{H}_{A}$ of the non-interacting atomic electron (i.e. in the absence of interactions with the quantized electromagnetic field) can be written as
\begin{equation}
\mathcal{H}_A=\Psi^\dagger(\br)\left[-\frac{\hbar^2}{2m}\nabla^2+V(|\br-\mathbf{R}|)\right]\Psi(\br),\label{eqn:AtomicHam}
\end{equation}
where $\Psi(\br)$ is the Schr\"odinger field operator satisfying the anticommutation relation
\begin{equation}
\left\{\Psi(\br),\Psi^\dagger(\br)\right\}=\delta^{(3)}(\br-\br'),\label{eqn:SchroAntiComm}
\end{equation}
and $V(|\br-\mathbf{R}|)$ is the potential due to the immobile nucleus which we choose to be located well outside the dielectric (i.e. at least few Bohr radii away) at a position $\mathbf{R}$, so that there is no wave-function overlap between the atom and the solid. The atom is coupled to the electric field via its electric dipole moment. The Hamiltonian describing this coupling in the dipole approximation may be written as
\begin{equation}
H_{A-EM}=-\boldsymbol{\mu}\cdot\mathbf{E}(\mathbf{R}),
\end{equation}
i.e. we evaluate the electric field at the position of the nucleus. Here $\boldsymbol{\mu}$ is the electric-dipole moment operator which depends on the second-quantized fields $\Psi$ and $\Psi^\dagger$. It is convenient to expand the field operator $\Psi(\br)$ in terms of a complete set of atomic wave-functions satisfying
\begin{equation}
\left[-\frac{\hbar^2}{2m}\nabla^2+V(|\br-\mathbf{R}|)\right]\phi_n(\br)=E_n\phi_n(\br).
\end{equation}
If we write
\begin{equation}
\Psi(\br)=\sum_m c_m \phi_m(\br),\;\;\Psi^\dagger(\br)=\sum_m c^\dagger_m \phi^*_m(\br)\label{eqn:PsiExpansion}
\end{equation}
then it follows from Eq. (\ref{eqn:SchroAntiComm}) that the operators $c_m$ and $c_n^\dagger$ satisfy the equal-time anticommutation relation
\begin{equation}
\left\{c_n, c_m^\dagger\right\}=\delta_{mn}.\label{eqn:CComms}
\end{equation}
We use equations (\ref{eqn:PsiExpansion}) and (\ref{eqn:CComms}) to rewrite the Hamiltonians $H_A$ and $H_{EM}$ in a more useful form,
\begin{eqnarray}
H_{A}&=&\sum_n E_n c_n^\dagger c_n,\label{eqn:AtomicHamFinal}\\
 H_{A-EM}&=&-e\sum_{ij}c_i^\dagger c_j\langle i |\boldsymbol{\rho}|j\rangle\cdot\mathbf{E}(\mathbf{R}),\label{eqn:AtomicHamCouplingFinal}
\end{eqnarray}
where $-e\langle i |\boldsymbol{\rho}|j\rangle$ are the dipole matrix elements.

We shall follow the field theoretical approach of \cite{SmallBody} to calculate the energy-level shifts and modified spontaneous decay rates. In order to do so we need to locate the poles of the atomic propagator, which, once interactions have been switched on, are accessible only perturbatively. For these perturbative calculations we need to work in the interaction picture where the general expression for the perturbative expansion of a Green's function of the field $\Psi$ under the influence of the interaction $H_{\rm I}$ is \cite{Fetter}
\begin{eqnarray}
\mathcal{G}(\br,\br',t,t')=\sum_{n=0}^\infty\left(-\frac{i}{\hbar}\right)^{n+1}\int\rd t_1\ldots\int\rd t_n   \hspace{1.4cm}\nonumber\\
\times\left\langle\Omega|\; {\T} \left[{\Psi}(\br,t){\Psi}^\dagger(\br',t'){H}_{\rm I}(t_1)\dots {H}_{\rm I}(t_n)\right]|\Omega\right\rangle_{\rm conn}.\;\;\;\label{eqn:expansion}
\end{eqnarray}
$\Psi$ is now the field operator in the Heisenberg picture and $|\Omega\rangle$ is the exact ground state of the non-interacting system. The subscript 'conn' indicates that only connected diagrams contribute, as disconnected diagrams drop out in the normalization of $|\Omega\rangle$.

Wick's theorem states that the terms appearing in the expansion (\ref{eqn:expansion}), when written out explicitly for a specific interaction Hamiltonian, turn out to be given entirely in terms of the propagators of the non-interacting fields.
We shall proceed by determining the non-interacting propagators of the atom, the polarization field, the bath, and the electromagnetic field. Then the interaction of the polarization field with the reservoir shall be treated exactly to all orders. Once this is accomplished the correction to the electromagnetic-field propagator caused by the presence of the absorptive dielectric can be calculated, which shall also be done exactly to all orders. This is going to give the dressed photon propagator that enters the final perturbative expansion of the atomic propagator whose poles give the energy-level shifts and changes in the transition rates.

\section{Unperturbed Feynman propagators}
\subsection{Atomic-electron propagator}
The unperturbed atomic-electron propagator corresponding to the Hamiltonian (\ref{eqn:AtomicHam}) or equivalently (\ref{eqn:AtomicHamFinal}) is defined as the time-ordered expectation value
\begin{equation}
G^{(0)}(\br,\br',t,t')=-\frac{i}{\hbar}\langle\Omega|\;\T\left[\Psi(\br,t)\Psi^\dagger(\br',t')\right]|\Omega\rangle,\label{eqn:AtomicPropDer1}
\end{equation}
where $\Psi$ is the Schr\"odinger field operator in the Heisenberg picture and $|\Omega\rangle$ is the exact ground state of the non-interacting system. We substitute the field operators written in terms of the atomic eigenfunctions, Eq. (\ref{eqn:PsiExpansion}), while remembering that we are in the Heisenberg picture where the operators $c_l$ and $c_m^\dagger$ are time-dependent, and find
\begin{equation}
G^{(0)}(\br,\br',t,t')=\sum_{l,m}\phi_l(\br)\phi^*_m(\br')G^{(0)}_{lm}(t,t')\label{eqn:AtomicPropDer2}
\end{equation}
with
\begin{equation}
G^{(0)}_{lm}(t,t')=-\frac{i}{\hbar}\langle\Omega|\;\T\left[c_l(t)c_m^\dagger(t')\right]|\Omega\rangle.\label{eqn:AtomicPropDer3}
\end{equation}
The time-dependence of the fermionic annihilation and creation operators is governed by the Hamiltonian (\ref{eqn:AtomicHamFinal}),
\begin{equation}
c_m(t)=c_m(0)e^{-iE_m t/\hbar},\;c^\dagger_m(t)=c^\dagger_m(0)e^{iE_m t/\hbar}.\label{eqn:TimeDemendenceofC}
\end{equation}
With that we can determine $G^{(0)}_{lm}(t,t')$ and obtain
\begin{equation}
G_{lm}^{(0)}(t-t')=-\frac{i}{\hbar}\theta(t-t')e^{-iE_l(t-t')/\hbar}\delta_{lm},
\end{equation}
where we have used the definition of the time-ordering operator and the fact that the vacuum state $|\Omega\rangle$ is annihilated by $c_m(0)$. 
Since $G_{lm}^{(0)}(t,t')$ is in fact dependent only on the time difference $t-t'$, we can work with its Fourier transform with respect to $t-t'\equiv\tau$
\begin{equation}
G_{lm}^{(0)}(E)=\int_{-\infty}^\infty\rd\tau\; e^{iE\tau/\hbar}\,G_{lm}^{(0)}(\tau)=\frac{1}{E-E_l+i\eta}\,\delta_{lm}.
\label{eqn:AtomicPropE}
\end{equation}
With this convention of Fourier transformation the $i\eta$-prescription ensures the correct causal behaviour of the propagator and guarantees the convergence of the integrals.

\subsection{Photon propagator}\label{sec:FreeSpacePhotProp}
To calculate the zeroth-order propagator of the displacement field $\mathbf{D}(\br,t)$ whose dynamics is governed by the Hamiltonian (\ref{eqn:EMHam}), which we emphasize does not include the coupling term (\ref{eqn:PolEMCoupling}), we note that the Heisenberg equations of motion imply
\begin{eqnarray}
\frac{\partial D_i(\br,t)}{\partial t}&=&\frac{1}{\mu_0}\epsilon^{ijk}\;\nabla_jB_k(\br,t),\label{eqn:Ddot}\\
\frac{\partial B_i(\br,t)}{\partial t}&=&-\frac{1}{\epsilon_0}\epsilon^{ijk}\;\nabla_jD_k(\br,t),\label{eqn:Bdot}
\end{eqnarray}
where $\epsilon^{ijk}$ is the Levi-Civita symbol and the sum over doubly occurring Cartesian indices is implied. Thus the displacement field $\mathbf{D}(\br,t)$ satisfies the homogeneous wave equation
\begin{equation}
(\nabla_i\nabla_j-\delta_{ij}\nabla^2)D_j(\br,t)+\mu_0\epsilon_0\frac{\partial^2}{\partial t^2}D_i(\br,t)=0.\label{eqn:WaveEqnForFreeDProp}
\end{equation}
The formal definition of the photon propagator reads:
\begin{equation}
D_{ij}(\br,\br',t,t')=-\frac{i}{\hbar}\langle 0 |\;\T\left[D_i(\br,t)D_j(\br',t')\right]|0\rangle,\label{eqn:PhotonDef}
\end{equation}
where $D_i(\br,t)$ is the displacement field operator in the Heisenberg picture and $|0\rangle$ is the exact ground state of the non-interacting electromagnetic field. We proceed by applying the differential wave-operator that appears in (\ref{eqn:WaveEqnForFreeDProp}) to this definition of the propagator, but we need to take care when applying the time derivative to a time-ordered product and observe that
\begin{eqnarray}
\frac{\partial}{\partial t} {\T}[A(t)B(t')]=\delta(t-t')[A(t),B(t)]+{\T} \left[ \frac{\partial A(t)}{\partial t} B(t') \right].\nonumber
\end{eqnarray}
Thus we find that the displacement field propagator $D^{(0)}_{ij}(\br,\br',t,t')$ satisfies the following differential equation: 
\begin{eqnarray}
\left(\nabla_i\nabla_j-\delta_{ij}\nabla^2+\mu_0\epsilon_0\frac{\partial^2}{\partial t^2}\right)D^{(0)}_{jk}(\br-\br',t-t')\nonumber\\
=\frac{\epsilon_0}{(2\pi)^3}\delta(t-t')\int \rd^3\bq(q_i q_k-\delta_{ik}\bq^2)e^{i\bq\cdot(\br-\br')},\label{eqn:FreePhotonDiffA}
\end{eqnarray}
where we have used the commutator
\begin{equation}
\left[\frac{\partial D_i(\br,t)}{\partial t},D_k(\br',t)\right]=\frac{i\hbar}{\mu_0}\left(\nabla_i\nabla_k-\delta_{ik}\nabla^2\right)\delta^{(3)}(\br-\br')\nonumber
\end{equation}
and the fact that spatial derivatives commute with time-ordering operator. From Eq.~(\ref{eqn:FreePhotonDiffA}) it is clear that the free-space photon propagator is translation-invariant in space and time, i.e. it depends only on the differences $\br-\br'$ and $\tau=t-t'$. Therefore one can find the solution of the differential equation through Fourier transformation. First we note that Maxwell's equation (\ref{eqn:Ddot}) implies that the displacement field is transverse, so that its propagator satisfies
\begin{equation}
\nabla_i D^{(0)}_{jk}(\br-\br',t-t')=0.
\end{equation}
Introducing the Fourier transform of the propagator
\begin{eqnarray}
D_{ik}^{(0)}(\bq,\w)=\int\rd^3(\br-\br')e^{-i\bq\cdot(\br-\br')} \hspace{2 cm}\nonumber\\
\times\int_{-\infty}^\infty \rd (t-t') e^{i\w(t-t')} D^{(0)}_{ik}(\br-\br',t-t')\;\;
\end{eqnarray}
we readily obtain its spectral representation
\begin{equation}
D_{ik}^{(0)}(\bq,\w)=\epsilon_0\frac{\delta_{ik}\bq^2-q_i q_k}{\w^2-\bq^2+i\eta}.
\label{eqn:PhotonFreeProp}
\end{equation}
We have displaced the poles in the denominator by $i\eta$ so that $D^{(0)}_{jk}(\br-\br',t-t')$ has the appropriate causality properties of a Feynman propagator.

\subsection{Polarization field propagator}\label{sec:FreePolProp}
The Hamiltonian density (\ref{eqn:PolarizationHam2}) describes a collection of mutually independent harmonic oscillators. The fact that the harmonic oscillator at $\br$ is unaffected by the oscillator at $\br+d\br$ allows us to introduce creation and annihilation operators,  $\mathbf{b}^\dagger(\br)$ and $\mathbf{b}(\br)$, for each harmonic oscillator,
\begin{eqnarray}
\mathbf{X}(\br)&=&\sqrt{\frac{\hbar}{2\mathcal{M}\tilde{\w}_{\rm T}}}\left[\mathbf{b}^\dagger(\br)+\mathbf{b}(\br)\right],\nonumber\\
\mathbf{P}(\br)&=&i\sqrt{\frac{\hbar\mathcal{M}\tilde{\w}_{\rm T}}{2}}\left[\mathbf{b}^\dagger(\br)-\mathbf{b}(\br)\right].\label{eqn:Xb}
\end{eqnarray}
The operators $\mathbf{b}^\dagger(\br)$ and $\mathbf{b}(\br)$ satisfy the equal-time commutation relations
\begin{equation}
\left[b_i(\br),b_j^\dagger(\br)\right]=\delta_{ij}\delta^{(3)}(\br-\br')\label{eqn:BComms}
\end{equation}
which follow directly from their definition and Eq.~(\ref{eqn:PolComm}). The operator $b_i(\br)$ annihilates the ground state of the oscillation in the $i$-th direction at $\br$. Using this property together with the commutation relation (\ref{eqn:BComms}), we can directly evaluate the polarization field propagator defined as
\begin{equation}
K_{ij}^{(0)}(\br,\br',t,t')=-\frac{i}{\hbar}\langle \Omega |T\left[X_i(\br,t)X_j(\br',t')\right]|\Omega\rangle .\label{eqn:PolarizationDef}
\end{equation}
Here $X_i(\br,t)$ is the polarization field operator in the Heisenberg picture and $|\Omega\rangle$ is the exact ground state of the non-interacting polarization field. When written in terms of the creation and annihilation operators, the Hamiltonian density (\ref{eqn:PolarizationHam2}) is of course diagonal in $\mathbf{b}^\dagger(\br)$ and $\mathbf{b}(\br)$ so that the time dependence of the creation and annihilation operators is harmonic,
\begin{equation}
\mathbf{b}(\br,t)=\mathbf{b}(\br,0)e^{-i\tilde{\w}_{\rm T}t},\;\mathbf{b}^\dagger(\br,t)=\mathbf{b}^\dagger(\br,0)e^{i\tilde{\w}_{\rm T}t}.
\end{equation}
We substitute the polarization field operators (\ref{eqn:Xb}) expressed in terms of the ladder operators into Eq. (\ref{eqn:PolarizationDef}) and observe that, due to the orthogonality of states, only terms proportional to $b_i b_i^\dagger$ contribute. Taking care of the appropriate time ordering of operators and using the commutator (\ref{eqn:BComms}) to move any annihilation operators to the right of creation operators, so that they act on the vacuum state $|\Omega\rangle$, we readily obtain
\begin{eqnarray}
K^{(0)}_{ij}(\br-\br',t-t')=-\frac{i}{2\mathcal{M}\tilde{\w}_{\rm T}}\delta_{ij}\delta^{(3)}(\br-\br')\hspace{1 cm}\nonumber\\
\times\left[\theta(t-t')e^{-i\tilde{\w}_{\rm T}(t-t')}+\theta(t'-t)e^{+i\tilde{\w}_{\rm T}(t-t')}\right],\;\;\;\label{eqn:PolarizationProp2}
\end{eqnarray}
with the frequency $\tilde{\w}_{\rm T}$ as defined in Eq. (\ref{eqn:FrequencyShift2}).
We shall need the Fourier transform of the polarization propagator with respect to the time difference $t-t'$
\begin{equation}
K^{(0)}_{ij}(\br-\br';\w)=\int_{-\infty}^\infty\rd(t-t')e^{i\w(t-t')}K^{(0)}_{ij}(\br-\br',t-t'),
\end{equation}
which is easily obtained from Eq. (\ref{eqn:PolarizationProp2}) and reads
\begin{equation}
K^{(0)}_{ij}(\br-\br';\w)=\frac{1}{\mathcal{M}}\frac{1}{\w^2-\tilde{\w}_{\rm T}^2+i\eta}\delta_{ij}\delta^{(3)}(\br-\br').\label{eqn:PolarizationProp}
\end{equation}
Since the polarization field operators are mutually independent there is no need for any special consideration of the boundaries of the dielectric medium at this stage. The boundaries are taken into account through the coupling function $g(\br)$, Eq. (\ref{eqn:Gfun}), and the field equation for the electromagnetic field, once coupled to the polarization field, will include the physical processes of reflection and refraction as it should. It is worth pointing out that an artificial restriction of the free polarization field to just the interior of the dielectric would lead to a much more complicated free propagator thereby causing unnecessary technical complications while not describing any different physics.

\subsection{Reservoir propagator}
The dynamics of the non-interacting reservoir field is governed by the Hamiltonian (\ref{eqn:BathHam}) which describes a set of independent harmonic oscillators. The propagator for the free reservoir field can be obtained by repeating the same steps as for the derivation of the propagator for the free polarization field in Section \ref{sec:FreePolProp}. Therefore, we do not repeat the details of the derivation but simply point out the similarity of the structure of the result to Eqs.~(\ref{eqn:PolarizationProp2}) and (\ref{eqn:PolarizationProp}). In the time domain the reservoir propagator reads
\begin{eqnarray}
H^{(0)}_{ij}(\br-\br',t-t',\nu,\nu')=-\frac{i}{2\rho_\nu\nu}\delta_{ij}\delta^{(3)}(\br-\br')\delta(\nu-\nu')\nonumber\\
\times\left[\theta(t-t')e^{-i\nu(t-t')}+\theta(t'-t)e^{+i\nu(t-t')}\right].\hspace*{9mm}
\end{eqnarray}
Its Fourier transform with respect to $t-t'$ is given by
\begin{eqnarray}
&&\hspace*{-6mm} H^{(0)}_{ij}(\br-\br',\nu,\nu';\w)\nonumber\\
&&=\int_{-\infty}^\infty\rd(t-t')\;e^{i\w(t-t')}H^{(0)}_{ij}(\br-\br',t-t',\nu,\nu'),\nonumber\\
&&=\frac{1}{\rho_\nu}\frac{\delta_{ij}}{\w^2-\nu^2+i\eta}\delta^{(3)}(\br-\br')\delta(\nu-\nu').\label{eqn:ReservoirProp}
\end{eqnarray}

\section{\label{sect:level3}Dressed propagators}
Having gathered all the unperturbed propagators, we can proceed to work out the propagators for the coupled fields. We are going to use a diagrammatic approach
to illustrate the workings of perturbation theory, i.e. we represent each term of the perturbative expansion (\ref{eqn:expansion}) by an appropriate Feynman diagram (c.f. e.g. \cite{Fetter}). To proceed with that, we need to lay down the Feynman rules for our approach. We have four different free propagators; accordingly, we associate with them four distinct lines:
\begin{figure}[h]
  \centering
    \includegraphics[width=7.5 cm, height=4.0 cm]{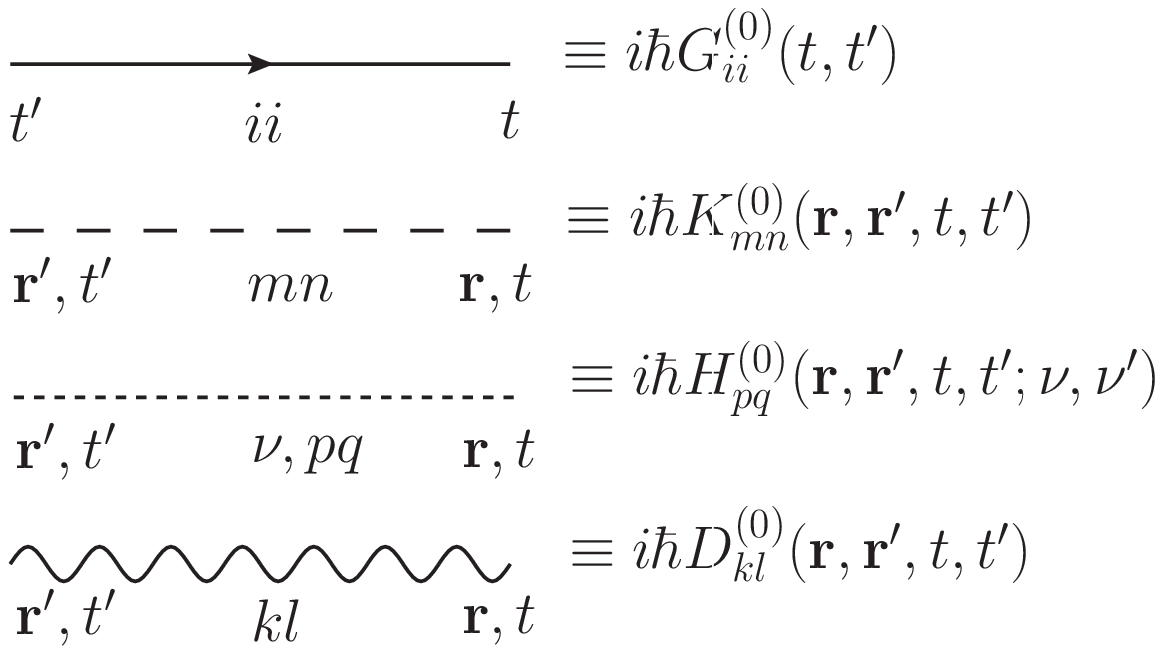}
  \label{fig:FeynmanRules}
\end{figure}

\noindent We shall need to consider three interaction Hamiltonians in turn, $H_{\rm P-R}$, $H_{\rm P-EM}$ and $H_{\rm A-EM}$:
\begin{eqnarray}
H_{\rm P-R}&=&-\int \rd^3\br\int_0^\infty\rd\nu\rho_\nu\nu^2\mathbf{X}(\br)\cdot \mathbf{Y}_\nu(\br),\label{eqn:Int1}\\
H_{\rm P-EM}&=&-\frac{1}{\epsilon_0}\int \rd^3\br g(\br)\mathbf{D}(\br)\cdot\mathbf{X}(\br),\label{eqn:Int2}\\
H_{A-EM}&=&-\frac{1}{\epsilon_0}\sum_{ij}c_i^\dagger c_j\;\boldsymbol{\mu}_{ij}\cdot \mathbf{D}(\mathbf{R}).\;\;\;\;\label{eqn:Int3}
\end{eqnarray}
Note that we have introduced the shorthand $\boldsymbol{\mu}_{ij}=\langle i |\boldsymbol{\mu}|j\rangle$ for the matrix elements of the atomic electric dipole moment operator $\boldsymbol{\mu}$.
These interaction Hamiltonians yield the following Feynman rules for the vertices between the lines defined above:
\begin{figure}[h]
  \centering
    \includegraphics[width=8.0 cm, height=5.4 cm]{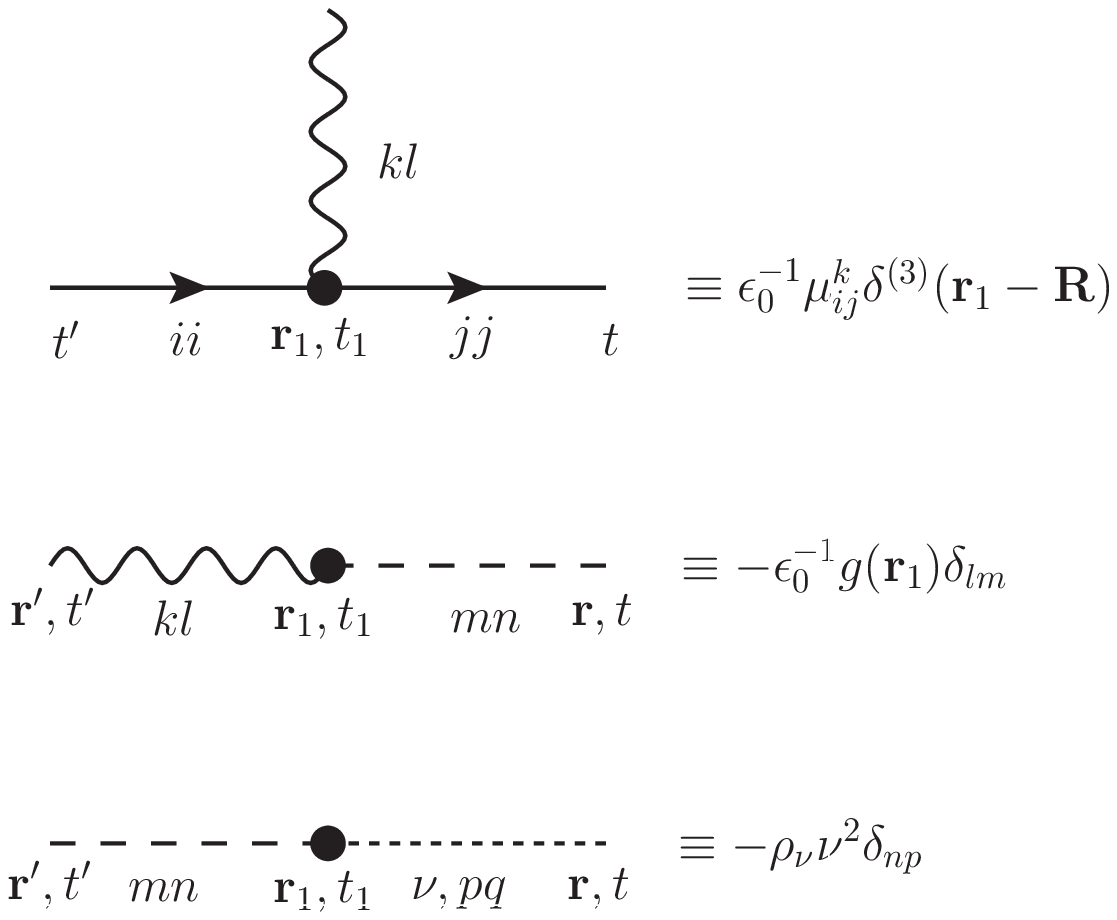}
  \label{fig:FeynmanRules2}
\end{figure}

\noindent To compute a diagram one has to sum over all internal indices and integrate over internal times, internal coordinates, and reservoir oscillator frequencies $\nu$ and $\nu'$. As mentioned earlier, the subscript 'conn' in Eq.~(\ref{eqn:expansion}) means that the summation in that equation runs only over those terms that correspond to connected Feynman diagrams. Furthermore, topologically equivalent diagrams, i.e. those that can be obtained by permuting the factors $H_{\rm I}(t_i)$ in Eq.~(\ref{eqn:expansion}), are counted only once, and therefore we have omitted the factor of $1/n!$ that would otherwise have arisen in a straightforward expansion of the time-ordered exponential of the interaction Hamiltonian in perturbation theory.  

\subsection{Dressing the polarization line}\label{sec:DressPol}
The polarization field interacts with the reservoir; all these interactions in their entirety "dress" the polarization field.
We choose to represent the dressed polarization propagator by a bold dashed line:
\begin{figure}[h]
  \centering
    \includegraphics[width=6 cm, height=1.0 cm]{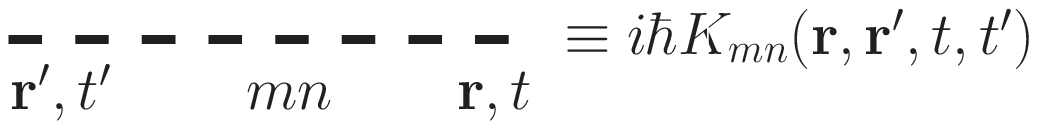}
  \label{fig:DressedPolRule}
\end{figure}

\noindent From the interaction Hamiltonian (\ref{eqn:Int1}) and the associated the Feynman rules one can see that the polarization line can only ever connect to exactly one reservoir line. Hence the complete set of all possible interactions corresponding to the expansion (\ref{eqn:expansion}) is represented by the following sequence of Feynman diagrams:
\begin{figure}[h]
  \centering
    \includegraphics[width=8.7 cm, height=2.0 cm]{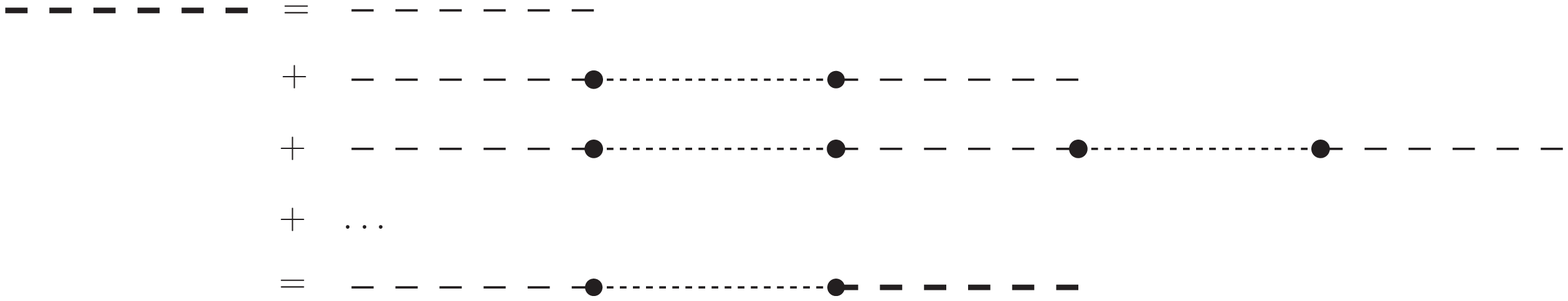}
  \label{fig:Dyson1}
\end{figure}

\noindent The equivalent analytical expression is the Dyson equation for the dressed polarization propagator; it reads
\begin{eqnarray}
K_{mn}(\br,\br',t,t')=K^{(0)}_{mn}(\br,\br',t,t')\hspace{3.5 cm}\nonumber\\
+\sum_{l,p}\int_{-\infty}^\infty\rd t_1\int_{-\infty}^\infty\rd t_2\int \rd^3 \br_1\int \rd^3 \br_2 \int_{0}^\infty\rd \nu \int_{0}^\infty\rd \nu'\hspace{1.0 cm}
\nonumber\\
\times K^{(0)}_{ml}(\br,\br_1,t,t_1)
H^{(0)}_{lp}(\br_1,\br_2,t_1,t_2,\nu,\nu')K_{pn}(\br_2,\br',t_2,t')\nonumber
\end{eqnarray}
Despite being an integral equation, the above equation is easily solved exactly. Upon substituting Eqs. (\ref{eqn:PolarizationProp}) and (\ref{eqn:ReservoirProp}) we can easily carry out the spatial integrations. Then we Fourier transform with respect to $t-t'$,
\begin{equation}
K_{mn}(\br,\br';\w)=\int_{-\infty}^\infty\rd(t-t')e^{i\w(t-t')}K_{mn}(\br,\br',t-t')\;, \nonumber
\end{equation}
and find the following expression for the dressed polarization field propagator in the frequency domain
\begin{eqnarray}
K_{mn}(\br-\br';\w)=K(\w)\delta^{(3)}(\br-\br')\delta_{mn}
\label{eqn:PolFinal}
\end{eqnarray}
with
\begin{equation}
K(\w)=\frac{1}{\mathcal{M}}\left[\w^2-\w_{\rm T}^2-\w_{\rm P}^2-\frac{\w^2}{\mathcal{M}}\int_0^\infty\rd\nu\frac{\rho_\nu\nu^2}{\w^2-\nu^2+i\eta}\right]^{-1}.
\label{eqn:Kdef}
\end{equation}
Note that $K(\w)$ is an even function of $\w$. The plasma frequency $\omega_{\rm P}$ was defined below Eq.~(\ref{eqn:FrequencyShift2}).

\subsection{Dressing the photon line}
\label{sec:PhotonDress}
The coupling (\ref{eqn:Int2}) between the dressed polarization field and the electromagnetic field has formally the same form as the coupling (\ref{eqn:Int1}) between the bare polarization field and the reservoir. By analogy with the previous section, we write down the graphical equation for the dressed photon propagator as
\begin{figure}[h]
  \centering
    \includegraphics[width=8.5 cm, height=.50 cm]{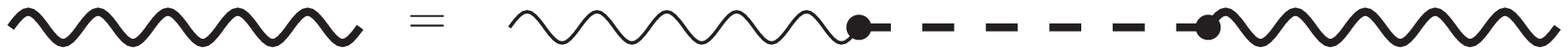}
  \label{fig:Dyson2}
\end{figure}

\noindent
where the bold wavy line denotes the dressed photon propagator i.e.
\begin{figure}[h]
  \centering
    \includegraphics[width=6 cm, height=1.0 cm]{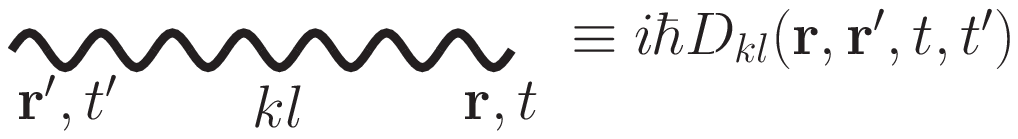}
  \label{fig:DressedPhotRule}
\end{figure}

\noindent
The corresponding analytical expression reads
\begin{eqnarray}
D_{ik}(\br,\br',t,t')=D^{(0)}_{ik}(\br-\br',t-t')\hspace{2.9 cm}\nonumber\\
+\frac{1}{\epsilon_0^2}\sum_{j,l}\int_{-\infty}^\infty\rd t_1\int_{-\infty}^\infty\rd t_2\int \rd^3 \br_1\int \rd^3 \br_2\;g(\br_1)g(\br_2)\hspace{1 cm}\nonumber\\
\times D_{ij}^{(0)}(\br-\br_1,t-t_1)
K_{jl}(\br_1-\br_2,t_1,t_2) D_{lk}(\br_2,\br',t_2,t').\nonumber\\\label{eqn:DysonForPhotons}
\end{eqnarray}
Now recall the discussion following Eq. (\ref{eqn:FrequencyShift1}) of the shifted eigenfrequency  $\tilde{\w}_{\rm T}$ of the polarization field. It enters the Dyson equation (\ref{eqn:DysonForPhotons}) through the dressed polarization field propagator $ K_{jl}(\br_1-\br_2,t_1,t_2)$. As we noted earlier, according to Eq. (\ref{eqn:FrequencyShift1}), the shifted eigenfrequency  $\tilde{\w}_{\rm T}$ suddenly jumps at the boundary of the region where the polarization field interacts with the electromagnetic field i.e. where the coupling function $g(\br)=1$. However, it is now apparent that this discontinuity is unproblematic because all spatial integrations in Eq.~(\ref{eqn:DysonForPhotons}) are limited to the region of space where $g(\br)=1$.

To simplify Eq.~(\ref{eqn:DysonForPhotons}) we note that one of the spatial integrations is trivial due to the $\delta$ function in the dressed polarization field propagator (\ref{eqn:PolFinal}). Fourier transforming with respect to the time difference $t-t'$,
\begin{equation}
D_{ik}(\br,\br';\w)=\int_{-\infty}^\infty \rd(t-t')e^{i\w(t-t')}D_{ik}(\br,\br',t,t')\;,
\end{equation}
we find the Dyson equation for the dressed photon propagator in the frequency domain:
\begin{eqnarray}
D_{ik}(\br,\br';\w) = D^{(0)}_{ik}(\br-\br';\w)\hspace{3.5 cm}\nonumber\\
 + \frac{K(\w)}{\epsilon_0^2}\int \rd^3 \br_1 g(\br_1)D^{(0)}_{ij}(\br-\br_1;\w)D_{jk}(\br_1,\br';\w).\;\;\;\;
\label{eqn:Integral}
\end{eqnarray}
Here $K(\omega)$ is a complex-valued function of frequency that has originated from the dressed polarization field propagator and is given in (\ref{eqn:Kdef}); it will be shown to be related to the dielectric permittivity. Note that the dimensionless coupling function $g(\br)$ describing the geometry of the dielectric medium, as defined in Eq.~(\ref{eqn:Gfun}), is the only way the geometry enters in the calculation, by effectively defining the limits of the spatial integration in Eq. (\ref{eqn:Integral}). $D^{(0)}_{ik}(\br-\br',\w)$ is the free-space photon propagator in coordinate representation i.e. the inverse Fourier transform of Eq.~(\ref{eqn:PhotonFreeProp}),
\begin{equation}
D^{(0)}_{ik}(\br-\br',\w)=\frac{\epsilon_0}{(2\pi)^3}\int\rd^3\bq \;e^{i\bq(\br-\br')}\frac{\delta_{ik}\bq^2-q_i q_k}{\w^2-\bq^2+i\eta}.\label{eqn:FreeSpacePosRep}
\end{equation}
The solution of the integral equation (\ref{eqn:Integral}) is much less trivial than that of the equivalent equation for dressing the polarization line in Section \ref{sec:DressPol}. This is because translation invariance is lost when an inhomogeneous dielectric is introduced into the system. Here we report two ways of tackling the problem. First, we demonstrate that it is possible to solve the integral equation (\ref{eqn:Integral}) by direct iteration. The iteration method that we shall employ is inspired by Ref.~\cite{Valeri}. In order to explain it, we write Eq.~(\ref{eqn:Integral}) symbolically as
\begin{equation}
D=D^0+K D^0 \otimes D.
\end{equation}
Iteration of this equation yields the expansion
\begin{equation}
D=D^0+K D^0 \otimes D^0 + K^2 D^0 \otimes D^0 \otimes D^0 +\ldots,\label{eqn:expansionDs}
\end{equation}
which proves especially useful if the action of the operator $\mathcal{O}=D^0 \otimes$ on the free-space propagator $D^0$ amounts to a simple multiplication, i.e. if
\begin{equation}
\mathcal{O}D^0=D^0 \otimes D^0= C D^0, \label{eqn:Multiplication}
\end{equation}
where $C$ is some constant. Then Eq.~(\ref{eqn:expansionDs}) becomes a geometrical series
\begin{equation}
D=D^0\left(1+KC+K^2C^2+K^3C^3+\ldots\right)\;,\label{eqn:geometric_series}
\end{equation}
which we know how to sum to all orders.

An alternative approach, which we sketch in Appendix \ref{App:DressProp}, consists of converting the integral equation (\ref{eqn:Integral}) to a differential equation supplemented by Maxwell boundary conditions. In addition, in Appendix \ref{App:DressPropPhen}, for comparison with other theories, we construct the photon propagator using yet another, completely different method based on the phenomenological noise-current approach of Ref.~\cite{PhenQED}.

\begin{figure}[h]
  \centering
    \includegraphics[width=8.5 cm, height=4.5 cm]{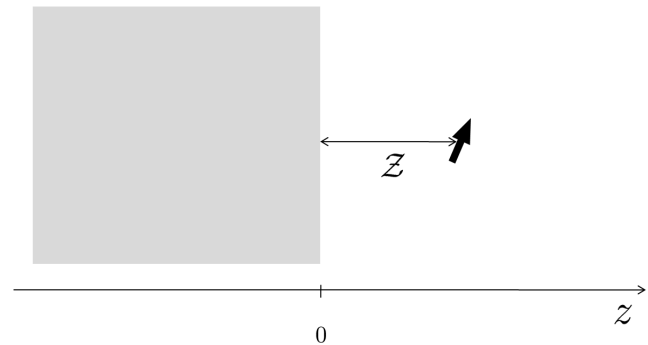}
  \caption{The atomic dipole is located a distance $\mathcal{Z}$ away from the dielectric half-space of complex and frequency-dependent permittivity $\epsilon(\w)$. The transverse propagator $D_{ik}(\br,\br';\w)$  of the dielectric displacement field in this geometry is given by Eq. (\ref{eqn:HSPropFinal}).}
  \label{fig:hspace}
\end{figure}

Let us now concentrate on the example geometry of a dielectric half-space occupying the $z<0$ region of space, cf. Fig. \ref{fig:hspace}, for which the coupling function $g(\br)$ in Eq. (\ref{eqn:Integral}) becomes $g(\br)=\theta(-z)$. Due to the boundary the problem has lost translation invariance in the $z$-direction, but not in directions parallel to the surface. In other words, the propagator depends only on the difference \mbox{$\brp-\brp'$}, but separately on $z$ and $z'$. It is convenient to work with quantities that have been Fourier transformed with respect to \mbox{$\brp-\brp'$}; e.g. for the dressed photon propagator we have
\begin{equation}
D_{ij}(z,z')=\int\rd^2(\brp-\brp')e^{-i\bqp\cdot(\brp-\brp')}D_{ij}(\brp-\brp',z,z'),\label{eqn:QZSpace}
\end{equation}
where for notational convenience we have suppressed the dependence on $\bqp$ and $\w$. Once Fourier transformed with respect to \mbox{$\brp-\brp'$}, the integral equation (\ref{eqn:Integral}) becomes
\begin{eqnarray}
D_{ik}(z,z')=D^{(0)}_{ik}(z-z')\hspace*{48mm}
\nonumber\\
+\frac{K(\w)}{\epsilon_0^2}\int_{-\infty}^0 \rd z_1 D^{(0)}_{ij}(z-z_1)D_{jk}(z_1,z').\hspace*{6mm}
\label{eqn:Halfspace1}
\end{eqnarray}
As is easily checked, this does not lend itself to iteration as it stands. Following Ref.~\cite{Valeri} we introduce an additional integral equation in order to enable the iteration process:
\begin{eqnarray}
D_{ik}(z,z')=D^{(\epsilon)}_{ik}(z-z')\hspace*{48mm}\nonumber\\
-\frac{K(\w)}{\epsilon_0^2}\int_{0}^\infty \rd z_1 D^{(\epsilon)}_{ij}(z-z_1)D_{jk}(z_1,z').\hspace*{6mm}
\label{eqn:Halfspace2}
\end{eqnarray}
Here $D^{(\epsilon)}_{ik}(z-z')$ is the Fourier-transformed photon propagator in a bulk medium, i.e. the solution of Eq.~(\ref{eqn:Integral}) with $g(\br)=1$. In order to justify Eq.~(\ref{eqn:Halfspace2}) let us recall that the part of the Hamiltonian density that describes the interaction of the photon field with the polarization field has the form
\begin{equation}
\mathcal{H}=\mathcal{H}_0-\frac{\theta(-z)}{\epsilon_0}\mathbf{X}(\br)\cdot\mathbf{D}(\br),
\end{equation}
where $\mathcal{H}_0$ is the Hamiltonian density of the non-interacting electromagnetic field. Using the fact that $\theta(-z)+\theta(z)=1$ we can also write
\begin{equation}
\mathcal{H}=\mathcal{H}_\epsilon+\frac{\theta(z)}{\epsilon_0}\mathbf{X}(\br)\cdot\mathbf{D}(\br),\label{eqn:H2}
\end{equation}
where $\mathcal{H}_\epsilon=\mathcal{H}_0-\mathbf{X}(\br)\cdot\mathbf{D}(\br)/\epsilon_0$ is the Hamiltonian density of the electromagnetic field interacting with an unbounded dielectric. Therefore, we have a choice: we can either correct the free-space photon propagator for the presence of the dielectric half-space or, equivalently, correct the photon propagator in a bulk dielectric for the absence of the medium in the other half-space. In other words, it is entirely up to us which Hamiltonian we take as the zeroth-order (exactly solvable) Hamiltonian when applying perturbation theory. The integral equation (\ref{eqn:Halfspace2}) corresponds to treating the electromagnetic field interacting with bulk medium as the zeroth-order, solved part of the problem. 

To proceed we need to find $D^{(0)}_{ik}(z-z')$ and $D^{(\epsilon)}_{ik}(z-z')$ appearing in Eq. (\ref{eqn:Halfspace1}) and (\ref{eqn:Halfspace2}), which can, in fact, be read off from the representations of these propagators as two-dimensional integrals over the momenta parallel to the surface. To find $D^{(0)}_{ik}(z-z')$ we carry out the $q_z$ integral in Eq.~(\ref{eqn:FreeSpacePosRep}) using the residue theorem. The result, written in a compact form, is
\begin{eqnarray}
D_{ik}^{(0)}(\br-\br',\w)=-\frac{i\epsilon_0}{2(2\pi)^2}\left(\nabla_i\nabla_k-\delta_{ik}\nabla^2\right)\hspace*{5mm}\nonumber\\
\times\int\rd^2\bqp e^{i\bqp\cdot\left(\brp-\brp'\right)}\dfrac{e^{ik_z|z-z'|}}{k_z}\label{eqn:ZIntegratedProp}
\end{eqnarray}
where $k_z$ is the $z$-component of the wave vector in vacuum and is given by $k_z=\sqrt{\w^2-\bqp^2+i\eta}$. The square root is taken such that the imaginary part of $k_z$ is always positive. Eq.~(\ref{eqn:ZIntegratedProp}) shows that some components of the Fourier transform of the free-space photon propagator are singular when crossing the $z=z'$ plane.

For deriving the photon propagator $D^{(\epsilon)}_{ik}(\br-\br';\w)$ in a bulk medium, we set $g(\br)=1$ in Eq.~(\ref{eqn:Integral}) and then Fourier transform this equation with respect to $\br-\br'$,
\begin{eqnarray}
D^{(\epsilon)}_{ik}(\bq,\w)=\epsilon_0\;\frac{\delta_{ik}\bq^2-q_i q_k}{\w^2-\bq^2+i\eta}\hspace{2.5 cm}\nonumber\\
+\frac{K(\w)}{\epsilon_0}\frac{\delta_{ij}\bq^2-q_i q_j}{\w^2-\bq^2+i\eta}\;D^{(\epsilon)}_{jk}(\bq,\w).
\end{eqnarray}
This matrix equation becomes an algebraic one when one takes the transversality of the propagator, $q_jD^{(\epsilon)}_{jk}(\bq,\w)=0$, into account. 
The calculation is straightforward and in coordinate space we obtain
\begin{equation}
D_{ik}^{(\epsilon)}(\br-\br',\w)=\frac{\epsilon_0\xi(\w)}{(2\pi)^3}\int\rd^3\bq \frac{\delta_{ik}\bq^2-q_i q_k}{\xi(\w)\w^2-\bq^2}\;e^{i\bq(\br-\br')}.\label{eqn:BulkPropagator}
\end{equation}
Note that the function $\xi(\w)$ that appears in Eq.~(\ref{eqn:BulkPropagator}) should not be interpreted as the dielectric function of the bulk medium.
It is an even function of the frequency $\w$ and may be written explicitly as
\begin{eqnarray}
\xi(\w)&=&\left(1+\dfrac{K(\w)}{\epsilon_0}\right)^{-1} \label{eqn:XiDef}\\
&=&1+\dfrac{1}{\epsilon_0\mathcal{M}}\left[\w_{\rm T}^2-\w^2-\frac{\w^2}{\mathcal{M}}\int_0^\infty\rd\nu\frac{\rho_\nu\nu^2}{\nu^2-\w^2-i\eta}\right]^{-1},\nonumber
\end{eqnarray}
where we have used Eq. (\ref{eqn:Kdef}). Thus the $\w$-dependence of $\xi(\w)$ is not consistent with the causality requirements usually imposed on response functions, i.e.~with Kramers-Kronig relations. This is because we have calculated a Feynman propagator and not a retarded Green's function. The dielectric function of this model is discussed in Appendix \ref{App:Epsilon}. We derive the required Fourier representation of the propagator in the bulk medium by carrying out the $q_z$ integral in Eq.~(\ref{eqn:BulkPropagator}) and obtain
\begin{eqnarray}
D^{(\epsilon)}_{ij}(\br-\br',\w)=-\frac{i\epsilon_0\xi(\w)}{2(2\pi)^2}\left(\nabla_i\nabla_k-\delta_{ik}\nabla^2\right)\nonumber\\
\times\int\rd^2\bqp e^{i\bqp\cdot\left(\brp-\brp'\right)}\dfrac{e^{ik_{zd}|z-z'|}}{k_{zd}},
\label{eqn:ZIntegratedPropA}
\end{eqnarray}
in complete analogy with the formula for the free-space propagator, Eq. (\ref{eqn:ZIntegratedProp}). Here $k_{zd}=\sqrt{\xi(\w)\w^2-\bqp^2}$ is the $z$-component of the complex wave vector in the medium with an always positive imaginary part. 

We may now proceed by substituting Eq. (\ref{eqn:Halfspace1}) into Eq. (\ref{eqn:Halfspace2})
\begin{eqnarray}
D_{ik}(z,z')=D_{ik}^{(\epsilon)}(z-z')\hspace{5 cm}\nonumber\\
-\frac{K(\w)}{\epsilon_0^2}\int_{0}^{\infty} \rd z_1 D^{(\epsilon)}_{ij}(z-z_1)D^{(0)}_{jk}(z_1-z')\hspace{2cm}\nonumber\\
-\frac{K^2(\w)}{\epsilon_0^4}\int_0^\infty\rd z_1 \int_{-\infty}^0\rd z_2 D_{ij}^{(\epsilon)}(z-z_1)\hspace{2 cm}\nonumber\\
\times D_{jl}^{(0)}(z_1-z_2)D_{lk}(z_2,z')\hspace{1 cm}
\label{eqn:Combined}
\end{eqnarray}
and focusing our attention on the solution of the case $z<0$ and $z'>0$, i.e. when the source is located outside the dielectric and the observation point is inside the material. The solution for the case {$z,z'>0$} can then be obtained by applying the integral equation (\ref{eqn:Halfspace1}). The advantage of introducing Eq.~(\ref{eqn:Combined}) is that it facilitates iteration as it turns out that when $D_{ik}$ on the right-hand side is replaced by $D_{ik}^{(\epsilon)}$
the action of the double-integral operator in the last term reduces to a matrix multiplication,
\begin{eqnarray}
\int_0^\infty\rd z_1 \int_{-\infty}^0\rd z_2 D_{ij}^{(\epsilon)}(z-z_1)D_{jl}^{(0)}(z_1-z_2)D^{(\epsilon)}_{lk}(z_2-z')\nonumber\\
=C_{ij}D^{(\epsilon)}_{lk}(z-z'),\hspace{.5 cm}\label{eqn:ShowMltpy2}
\end{eqnarray} 
with the matrix $C_{ij}$ independent of $z$ and $z'$. In order to efficiently verify and make use of assertion (\ref{eqn:ShowMltpy2}) let us point out some useful facts. First we recall that
\begin{equation}
q^2\delta_{ik}-q_i q_k=\w^2\left[e_i^\TE(\bq)e_k^\TE(\bq)+e_i^\TM(\bq)e_k^\TM(\bq)\right],\label{eqn:Useful}
\end{equation} 
where $\bq=(\bqp,k_z)$ is the wave vector in vacuum and we have introduced the polarization vectors
\begin{eqnarray}
\mathbf{e}^{\TE}(\bqp)&=&\frac{1}{|\bqp|}(-q_y,q_x,0),\nonumber\\
\mathbf{e}^{\TM}(\bqp,k_z)&=&\frac{1}{|\bqp|\w}(q_xk_z,q_yk_z,-\bqp^2),\label{eqn:PolarizationVec}\\
\mathbf{e}^{\TM}(\bqp,k_{zd})&=&\frac{1}{|\bqp|\sqrt{\xi(\w)}\w}(q_xk_{zd},q_yk_{zd},-\bqp^2)\nonumber.
\end{eqnarray}
We have listed $\mathbf{e}^{\TM}(\bqp,k_{zd})$ explicitly to point out the additional factor of $\xi^{-1/2}(\w)$ in its normalization. In the following we will suppress the insignificant dependence of the polarization vectors on $\bqp$. Relation (\ref{eqn:Useful}) is simply a statement of the completeness property of the polarization vectors (\ref{eqn:PolarizationVec}), but it allows us to write
\begin{eqnarray}
&&\left(\nabla_i\nabla_k-\delta_{ik}\nabla^2\right)e^{i\bqp\cdot\left(\brp-\brp'\right)+ik_z|z-z'|}\nonumber\\
&&=\w^2e^{i\bqp\cdot\left(\brp-\brp'\right)}\sum_{\lambda}
\left\{
\begin{array}{lr}
e_i^\lambda(k_z)e_k^\lambda(k_z)e^{ik_z(z-z')}, & z>z' \\
e_i^\lambda(-k_z)e_k^\lambda(-k_z)e^{-ik_z(z-z')}, & z<z'
\end{array}
\right.\hspace*{5mm}
\end{eqnarray}
so that the partial Fourier transform of the free-space propagator (\ref{eqn:ZIntegratedProp}) may be written as
\begin{eqnarray}
&&D^{(0)}_{ij}(z-z')=
-\frac{i\epsilon_0\w^2}{2k_z}\sum_{\lambda}\nonumber\\
&&\hspace*{15mm}\times\left\{
\begin{array}{lr}
e_i^\lambda(k_z)e_k^\lambda(k_z)e^{ik_z(z-z')}, & z>z' \\
e_i^\lambda(-k_z)e_k^\lambda(-k_z)e^{-ik_z(z-z')}, & z<z'
\end{array}
\right. .\hspace*{5mm} \label{eqn:ZIntegratedProp1}
\end{eqnarray}
We emphasize that the above representation of the free-space propagator is not valid at the point $z=z'$, where the $z$-derivatives in Eq.~(\ref{eqn:ZIntegratedProp}) acting on $e^{ik_z|z-z'|}$ would produce additional terms proportional to a delta function.
Similarly we have
\begin{eqnarray}
&&D^{(\epsilon)}_{ij}(z-z')=-\frac{i\epsilon_0\xi^2(\w)\w^2}{2k_{zd}}\sum_{\lambda}\nonumber\\
&&\hspace*{15mm}\times\left\{
\begin{array}{lr}
e_i^\lambda(k_{zd})e_k^\lambda(k_{zd})e^{ik_{zd}(z-z')}, & z>z', \\
e_i^\lambda(-k_{zd})e_k^\lambda(-k_{zd})e^{-ik_{zd}(z-z')}, & z<z'.
\end{array}
\right..\hspace*{5mm}\label{eqn:ZIntegratedProp1A}
\end{eqnarray}
Eqs.~(\ref{eqn:ZIntegratedProp1}) and (\ref{eqn:ZIntegratedProp1A}) show that the free-space and bulk-medium propagators can be split into separate contributions from the transverse electric and transverse magnetic polarizations
\begin{equation}
D^{(...)}_{ij}(z-z')=\sum_\lambda D^{(...)}_{\lambda,ij}(z-z') .\label{eqn:Spliting}
\end{equation}
Most of the further calculations are very much simplified if one takes into account that scalar products of polarization vectors with different $z$-components are diagonal in the polarization indices, i.e. we have
\begin{equation}
e^\lambda_i(\bqp,q_z)e_i^\sigma(\bqp,p_z)=f^{\lambda}(q_z,p_z)\delta_{\lambda\sigma}.\label{eqn:PolProdDiag}
\end{equation}
The function $f$ is equal to 1 for the TE mode, and for the TM mode it reads
\begin{equation}
f^{\TM}(q_z,p_z)=\dfrac{q_zp_z+\bqp^2}{\sqrt{\bqp^2+q_z^2}\sqrt{\bqp^2+p_z^2}}\label{eqn:FDef}.
\end{equation}
This is very useful because it shows that not only a single propagator, as in Eq.~(\ref{eqn:Spliting}), but also a product of propagators can always be split into separate contributions from the transverse electric and transverse magnetic modes, i.e. we can always write 
\begin{eqnarray}
\ldots D_{ij}(z-z_1)D_{jl}(z_1-z_2)D_{lk}(z_2-z')\ldots \hspace{2.2 cm}\nonumber\\
=\sum_\lambda \ldots D_{\lambda,ij}(z-z_1)  D_{\lambda,jl}(z_1-z_2) D_{\lambda,lk}(z_2-z').\ldots \nonumber
\end{eqnarray}
This is true for an arbitrary number of propagators. 

We can now proceed to verifying Eq. (\ref{eqn:ShowMltpy2}). First we note that the arguments of all three propagators entering Eq. (\ref{eqn:ShowMltpy2}) have a definite sign. Indeed we have
\begin{equation}
z-z_1<0,\;\;z_1-z_2>0,\;\;z_2-z'<0.
\end{equation}
Thus, it follows from Eqs.~(\ref{eqn:ZIntegratedProp1}) and (\ref{eqn:ZIntegratedProp1A}) that the propagators entering the integral in Eq.~(\ref{eqn:ShowMltpy2}) are given by
\begin{eqnarray}
D^{(0)}_{ij}(z-z')&=&-\frac{i\epsilon_0\w^2}{2k_z}e^{ik_z(z-z')}\sum_\lambda e_i^\lambda(k_z)e^\lambda_j(k_z),\nonumber\\
D^{(\epsilon)}_{ij}(z-z')&=&-\frac{i\epsilon_0\xi^2(\w)\w^2}{2k_{zd}}e^{-ik_{zd}(z-z')}\nonumber\\
&\; &\hspace{1 cm}\times\sum_\lambda e_i^\lambda(-k_{zd})e^\lambda_j(-k_{zd}).
\end{eqnarray}
With this we can evaluate the integrals in Eq.~(\ref{eqn:ShowMltpy2}) and find that
\begin{eqnarray}
\int_0^\infty\rd z_1 \int_{-\infty}^0\rd z_2 D_{ij}^{(\epsilon)}(z-z_1)D_{jl}^{(0)}(z_1-z_2)D^{(\epsilon)}_{lk}(z_2-z')\nonumber\\
=\frac{\epsilon_0^4}{K^2(\w)}\sum_\lambda\dfrac{r_\lambda^2}{1-r_\lambda^2}D^{(\epsilon)}_{\lambda,lk}(z-z'),\hspace{.5 cm}
\label{eqn:Multply}
\end{eqnarray}
where we have used Eq.~(\ref{eqn:XiDef}). Here $r_\lambda$ is the Fresnel coefficient for reflection from a half-space. Since all the Fresnel coefficients for reflection and transmission at a half-space will be needed later on, we list them here:
\begin{eqnarray}
r_\TE=\frac{k_z-k_{zd}}{k_z+k_{zd}},\;\;\;r_\TM=\frac{\xi(\w)k_z-k_{zd}}{\xi(\w)k_z+k_{zd}},\nonumber\\
t_\TE=\frac{2k_z}{k_z+k_{zd}},\;\;\;t_\TM=\frac{2\sqrt{\xi(\w)}k_z}{\xi(\w)k_z+k_{zd}}.
\label{eqn:Fresnel}
\end{eqnarray}
The significance of Eq.~(\ref{eqn:Multply}) is that it allows us to iterate the integral equation (\ref{eqn:Combined}) along the lines of Eqs. (\ref{eqn:Multiplication}) and (\ref{eqn:geometric_series}). 
The iterative process is now straightforward and, thanks to relation (\ref{eqn:Multply}), leads to two separate geometric series for the two polarizations
\begin{eqnarray}
D_{\lambda,ij}(z,z') = \bigg[ D^{(\epsilon)}_{\lambda,ij}(z-z') \hspace{3.5 cm}\nonumber\\
-\frac{K(\w)}{\epsilon_0^2}\int_0^\infty\rd z_1  D^{(\epsilon)}_{\lambda,ij}(z-z_1)
 D_{\lambda,jk}^{(0)}(z_1-z') \bigg]\hspace{.8 cm}\nonumber\\
\times\left[1-\left(\dfrac{r_\lambda^2}{1-r_\lambda^2}\right)+\left(\dfrac{r_\lambda^2}{1-r_\lambda^2}\right)^2+\ldots\right]\hspace{.5 cm}\label{eqn:IntegralEqIterated2}
\end{eqnarray}
These geometric series can easily be summed up to all orders to give the exact photon propagator for the case $z<0,\; z'>0$. In order to cast the result into a familiar form, 
we explicitly evaluate the integral in the second line, which requires some care. The integral that needs to be evaluated is
\begin{equation}
I^\lambda_{ik}(z,z')=\frac{K(\w)}{\epsilon_0^2}\int_0^\infty\rd z_1 D^{(\epsilon)}_{\lambda,ij}(z-z_1) D_{\lambda,jk}^{(0)}(z_1-z').\label{eqn:I}
\end{equation}
Here the argument of $D^{(\epsilon)}$ is always negative, $z-z_1<0$, whereas the sign of $z_1-z'$ can be both positive and negative. Therefore we need to take into account that the propagator (\ref{eqn:ZIntegratedProp}) is discontinuous at $z_1=z'$ and contains singular terms proportional to $\delta(z_1-z')$. 

In order to correctly evaluate the integral (\ref{eqn:I}) we represent the differential operator in Eq.~(\ref{eqn:ZIntegratedProp}) using the polarization vectors written out in terms of derivatives. Using the completeness relation of the transverse polarization vectors we may symbolically write
\begin{equation}
\nabla_i\nabla_k-\delta_{ik}\nabla^2=-\nabla^2\sum_\lambda e^\lambda_i(\boldsymbol{\nabla})e^\lambda_k(\boldsymbol{\nabla}).
\end{equation}
With this, the propagators entering the integral (\ref{eqn:I}) are given by
\begin{eqnarray}
D^{(0)}_{\lambda,ij}(z_1-z') &=&-\frac{i\epsilon_0}{2k_z}\left(\bqp^2-\nabla^2_{z'}\right) e_i^\lambda(-\nabla_{z'})e^\lambda_j(-\nabla_{z'})\nonumber\\
&\;&\hspace{2 cm}\times e^{ik_z|z_1-z'|},\\
D^{(\epsilon)}_{\lambda,ij}(z-z_1) &=&-\frac{i\epsilon_0\xi^2(\w)\w^2}{2k_{zd}} e_i^\lambda(-k_{zd})e^\lambda_j(-k_{zd})\nonumber\\
&\;&\hspace{2 cm}\times e^{-ik_{zd}(z-z_1)}.
\end{eqnarray}
Note that in $D^{(0)}$ we have changed the $z$-derivatives to act on $z'$ rather than on $z$ so that they could be pulled outside the integral in (\ref{eqn:I}). Now it is straightforward to demonstrate that the integral (\ref{eqn:I}) is given by
\begin{eqnarray}
I^\lambda_{ik}(z,z')=D^{(\epsilon)}_{\lambda,ik}(z-z')\hspace{4.5 cm}
\nonumber\\
 -\frac{i\epsilon_0\xi(\w)\w^2}{2k_{zd}}\dfrac{1}{t_\lambda}e^\lambda_i(-k_{zd})e^\lambda_k(-k_{z})e^{-ik_{zd}z+ik_z z'}\;,\hspace{1 cm}
\end{eqnarray}
whose first term exactly cancels the bulk dielectric propagator in the first line of Eq. (\ref{eqn:IntegralEqIterated2}). The remaining term yields the final result
\begin{eqnarray}
D_{ij}(z,z')=-\frac{i\epsilon_0\w^2}{2k_z}\sum_\lambda\left[\xi(\w)e^\lambda_i(-k_{zd})e_j^\lambda(-k_z)t_\lambda\right]\nonumber\\
\times e^{-ik_{zd}z+ik_z z'},\hspace{.5 cm}\label{eqn:HSPropConfirmed}
\end{eqnarray}
with the transmission coefficient as given in (\ref{eqn:Fresnel}). This formula describes the vacuum-dielectric transmission i.e. it is valid for $z<0,\;z'>0$. It is a straightforward calculation to plug equation (\ref{eqn:HSPropConfirmed}) into (\ref{eqn:Halfspace1}) and obtain the photon propagator for the case $z,z'>0$. In the region $z'>0$ the final result for the dressed photon propagator Fourier transformed back to coordinate space may be written as:
\begin{widetext}
\begin{eqnarray}
D_{ij}(\br,\br';\w)=\theta(z)D^{(0)}_{ij}(\br-\br';\w)-\frac{i\epsilon_0}{(2\pi)^2}\sum_{\lambda}\int\rd^2\bqp \frac{\w^2}{2k_z} e^{i\bqp\cdot(\brp-\brp')}\hspace{6 cm}\nonumber\\
\times\bigg\{
\theta(-z)\left[\xi(\w)e^\lambda_i(\bqp,-k_{zd})e_j^\lambda(\bqp,-k_z)\;t_\lambda\right]e^{-ik_{zd}z+ik_z z'}
+\theta(z)\left[e^\lambda_i(\bqp,k_{z})e_j^\lambda(\bqp,-k_z)\;r_\lambda \right]e^{ik_{z}(z+z')}\bigg\}.\;\;\;\;\label{eqn:HSPropFinal}
\end{eqnarray}
\end{widetext}
In the calculations of the energy-level shifts of an atom placed outside an absorbing dielectric material, to be discussed in the following section, we shall need the propagator for the case $z,z'>0$. In that case Eq.~(\ref{eqn:HSPropFinal}) shows that the propagator splits into a free-space part $D^{(0)}_{ij}(\br-\br';\w)$, which is not interesting as it just yields the standard (position-independent) Lamb shift, and a correction due to reflection at the boundary, which we shall call $D^{(r)}_{ij}(\br,\br';\w)$ and which gives a rise to the position-dependent Casimir-Polder shift. As we shall treat the atom-field interaction in the dipole approximation, we are going to need the reflected part of the propagator $D^{(r)}_{ij}(\br,\br';\w)$ evaluated at equal arguments $\br=\br'=\mathbf{R}$, where $\mathbf{R}=(0,0,\mathcal{Z})$ is the position of the atom. In that case it simplifies considerably and can be written in the form
\begin{eqnarray}
\mathbf{D}^{(r)}(\mathcal{Z};\w)=-\frac{i\epsilon_0}{8\pi}\int_0^\infty\rd q_\| \frac{q_\|}{k_z}e^{2ik_z\mathcal{Z}}\hspace{3 cm}
\nonumber\\
\times
\left(
\begin{array}{ccc}
\w^2r_\TE-k_z^2r_\TM&0&0\\
0&\w^2r_\TE-k_z^2r_\TM&0\\
0&0&2q_\|^2r_R^\TM
\end{array}\right)\;\;\;\;\;\label{eqn:PhotonPropDiag}
\end{eqnarray}
with $k_z=\sqrt{\w^2-q_\|^2+i\eta}$, as before. Note that we have gone to polar coordinates, $q_x=q_\|\cos\phi,\;q_y=q_\|\sin\phi$, where the azimuthal integration annihilated the off-diagonal elements of equal-argument propagator $D^{(r)}_{ik}(\br,\br;\w)$.

As a final remark we would like to comment on the convergence of the series in Eq.~(\ref{eqn:IntegralEqIterated2}). It clearly converges provided
\begin{equation}
\left|\dfrac{r_\lambda^2}{1-r_\lambda^2}\right|<1.\label{eqn:ConvergenceCond}
\end{equation}
However, there does not seem to be a physical significance to this condition. That the result for the propagator can be extended by analytic continuation to wave vectors not satisfying the condition (\ref{eqn:ConvergenceCond}) can be shown by solving the corresponding boundary-value problem, the procedure for which we sketch in Appendix \ref{App:DressProp}. 

\section{\label{sec:level4}Atomic propagator and electron self-energy}\label{sec:AtomicPropExpan}
In order to investigate the perturbative expansion of the atomic propagator (\ref{eqn:expansion}), we use the expansion in terms of atomic eigenstates, Eq.~(\ref{eqn:PsiExpansion}), and then work with the atomic propagator in that basis. In analogy to Eqs.~(\ref{eqn:AtomicPropDer2}) and (\ref{eqn:AtomicPropDer3}) for the unperturbed propagator, we obtain
\begin{eqnarray}
G_{ii}(t,t')=\sum_{n=0}^\infty\left(-\frac{i}{\hbar}\right)^{n+1}\int\rd t_1\ldots\int\rd t_n\hspace{2.5cm}\nonumber\\
\times \left\langle\Omega\left| {\rm T} \left[c_i(t)c_i^\dagger(t')H_{\rm A-EM}(t_1)\dots H_{\rm A-EM}(t_n)\right]\right|\Omega\right\rangle_{\rm conn}.\label{eqn:Atomic}
\end{eqnarray}
By using Wick's theorem to evaluate the ground-state expectation value of the time-ordered product of operators, one easily sees that the zeroth-order term is a propagator for non-interacting system and that the first-order correction vanishes because it is not possible to contract all of the operators. Therefore the lowest-order non-vanishing perturbative contributions come from terms of order $e^2$. Diagrams to this order have two vertices and therefore include a pair of disconnected tadpole diagrams, which are irrelevant as they go away in the process of normalization, and the physically important self-energy diagram which contains all the information about the energy-level shifts and decay rates,

\noindent
\begin{figure}[h]
  \centering
    \includegraphics[width=7.5 cm, height=1.7 cm]{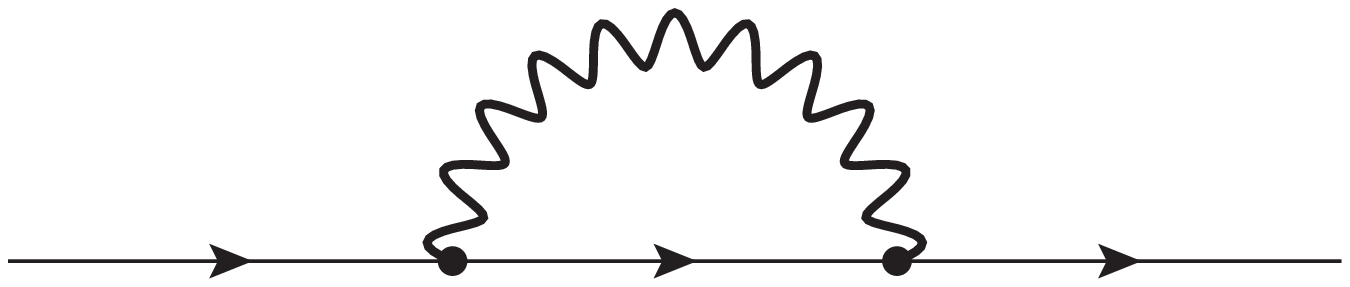}
  \label{fig:SelfEnergy}.
\end{figure}

\noindent
For calculating the perturbatively corrected, dressed atomic propagator it is in fact convenient to perform a partial summation and consider the following series of diagrams
\begin{figure}[h]
  \centering
    \includegraphics[width=8.5 cm, height=1.1 cm]{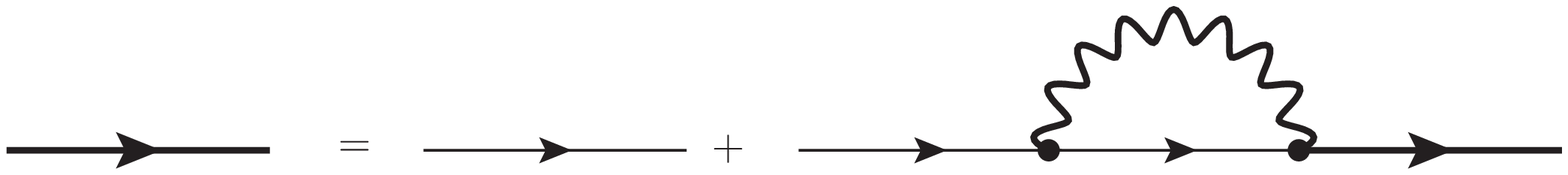}
  \label{fig:FullSelfEnergy}
\end{figure}

\noindent
where the thick solid line represents the dressed atomic propagator. The location of poles of the such constructed propagator is much more straightforward to work out than for the propagator with strictly only one-loop corrections. The above is a graphical representation of the Dyson equation, which expressed analytically reads
\begin{eqnarray}
G_{ii}(t,t')=G^{(0)}_{ii}(t,t')+\frac{i\hbar}{\epsilon_0^2}\sum_{k,l,m}\mu^k_{mi}\mu^l_{im}\int_{-\infty}^\infty\rd t_1\int_{-\infty}^\infty\rd t_2\nonumber\\
\times G_{ii}^{(0)}(t,t_1)G^{(0)}_{mm}(t_1,t_2)D_{kl}(\mathbf{R},\mathbf{R},t_1,t_2)G_{ii}(t_2,t').\hspace{.8 cm}
\label{eqn:DressedAtomic}
\end{eqnarray}
We note that here and in the following the indices $k$ and $l$ label just Cartesian components, but the sum over $m$ is a sum over intermediate atomic eigenstates, and $i$ is the atomic state whose energy shift we are seeking to determine. To this end we Fourier-transform (\ref{eqn:DressedAtomic}) with respect to $t-t'$, along the line of Eq.~(\ref{eqn:AtomicPropE}) and find the dressed atomic propagator
\begin{eqnarray}
G_{ii}(\mathcal{E})=\int_{-\infty}^{\infty}\rd(t-t')e^{i(t-t')\mathcal{E}/\hbar}G_{ii}(t-t')\nonumber\\
=\frac{1}{\mathcal{E}-E_i+i\eta-\Sigma_{ii}(\mathcal{E})}
\label{eqn:AtomicPoles}
\end{eqnarray}
with the self-energy insertion
\begin{equation}
\Sigma_{ii}(\mathcal{E})=
\frac{i\hbar}{2\pi\epsilon_0^2}\sum_{k,l,m}\mu_{mi}^k\mu_{im}^l\int_{-\infty}^\infty\rd\w\;\frac{D_{kl}(\mathbf{R},\mathbf{R};\w)}{\mathcal{E}-\hbar\w-E_m+i\eta}.
\label{eqn:SelfEnergy}
\end{equation}
The self-energy insertion (\ref{eqn:SelfEnergy}) contains the dressed photon propagator which in the case of an atom outside a dielectric half-space comprises:  \emph{(i)} the free-space photon propagator $D_{kl}^{(0)}(\mathbf{R},\mathbf{R};\w) $, which gives a rise to the position-independent Lamb shift, and \emph{(ii)} the reflected part $D_{kl}^{(r)}(\mathbf{R},\mathbf{R};\w) $, which yields the position-dependent Casimir-Polder shift. Thus the shift in the atomic energy-levels, given by the poles of Eq.~ (\ref{eqn:AtomicPoles}), 
can be written
\begin{equation}
\mathcal{E}-E_i=\Sigma^{(0)}_{ii}(\mathcal{E})+\Sigma^{(r)}_{ii}(\mathcal{E}).
\end{equation}
As we want to work out changes in the energy levels already corrected for the coupling between the atom and the free-space electromagnetic fields, we renormalize the energy-level shift by subtracting the self-energy associated with the free-space electromagnetic field $\Sigma^{(0)}_{ii}(\mathcal{E})$ and consider
\begin{equation}
\Delta \mathcal{E}^{\rm ren}_i\equiv \mathcal{E}-\bar{E}_i=\Sigma_{ii}^{(r)}(\mathcal{E}).
\label{eqn:RenResult}
\end{equation}
We use the symbol $\bar{E}_i\equiv E_i+\Sigma^{(0)}_{ii}(\mathcal{E})$ to represent the atomic energy levels already corrected for the free-space Lamb shift and decay rates.

Prima facie it may seem difficult to extract the energy shift from Eq.~(\ref{eqn:RenResult}) because it is an implicit equation whose right-hand side also depends on 
$\mathcal{E}$. However, if the energy shift we calculate is small compared to the difference in energy between the state under consideration and its nearest dipole-connected neighbours (which it needs to anyhow for perturbation theory to be applicable), then the shift can be extracted from Eq.~(\ref{eqn:RenResult}) by a single iteration leading to
\begin{eqnarray}
\Delta \mathcal{E}_i^{\rm ren}\approx\Sigma_{ii}^{(r)}(\bar{E}_i)\hspace{5 cm}\nonumber\\
=-\frac{i}{2\pi\epsilon_0^2}\sum_{k,l,m}\mu_{mi}^k\mu_{im}^l\int_{-\infty}^\infty\rd\w\;\frac{D^{(r)}_{kl}(\mathbf{R},\mathbf{R};\w)}{\w_{mi}+\w-i\eta}\hspace{.5 cm}
\label{eqn:ApproxPoles}
\end{eqnarray}
where we have abbreviated $\w_{mi}=\w_m-\w_i$. The $\w$-integral in (\ref{eqn:ApproxPoles}) can be restricted to the positive real axis by writing
\begin{equation}
\frac{1}{\w+\w_{mi}-i\eta}=\frac{\w-\w_{mi}}{\w^2-(\w_{mg}-i\eta)^2}
\end{equation}
and noting that $D^{(r)}_{kr}(\mathbf{R},\mathbf{R};\w)$ is even in $\w$ (see Section \ref{sec:PhotonDress} and Appendix \ref{App:Epsilon}). Then the term proportional to $\w$ is odd and vanishes when integrated over the real $\w$ axis. As the photon propagator is analytic in the first quadrant of the complex $\w$-plane, it is permissible, provided $\w_{mi}>0$, to rotate the contour of $\w$-integration by $\pi/2$ i.e. $\w\rightarrow i\w$. This applies when one considers an atom in its ground state. 

However, for an excited state $i$ of the atom one has $\w_{mi}<0$, which means that there will be poles in the first quadrant of the $\w$ plane due to the denominator in (\ref{eqn:ApproxPoles}). We would also to remark that the Fresnel reflection coefficients have poles in the complex plane at the location of trapped electromagnetic modes, which is not an issue in the case of a dielectric half-space but arises e.g.~for a dielectric slab \cite{slab} and other systems capable of wave-guiding \cite{Layered}.

We recall that $D^{(r)}_{kr}(\mathbf{R},\mathbf{R};\w)$ is diagonal, cf.~Eq.~(\ref{eqn:PhotonPropDiag}), and write down the final result for the energy shift in the form
\begin{equation}
 \Delta \mathcal{E}^{\rm ren}_i=\Delta \mathcal{E}_i+\Delta \mathcal{E}^{\star}_i
\end{equation}
with $\Delta \mathcal{E}_i$ and $\Delta \mathcal{E}^{\star}_i$ given by
\begin{eqnarray}
\Delta \mathcal{E}_i=
\frac{1}{\pi\epsilon_0^2}\sum_{k,m}|\mu_{im}^k|^2\int_{0}^\infty\rd\w\frac{\w_{mi}}{\w^2+\w^2_{mi}}D^{(r)}_{kk}(\mathbf{R},\mathbf{R};i\w)\nonumber\\ \label{eqn:ShiftFinal}\\
\Delta \mathcal{E}^{\star}_i =\frac{1}{\epsilon_0^2}\sum_{k,m}|\mu_{im}^k|^2 D^{(r)}_{kk}(\mathbf{R},\mathbf{R};|\w_{mi}|)\theta(-\w_{mi})\hspace{1 cm}\label{eqn:ShiftFinal1}
\end{eqnarray}
where $|\mu_{mi}^k|\equiv|\langle m|\mu^k|i\rangle|$ are the matrix elements of the $k$-th component of the electric dipole moment operator. The quantity $\Delta \mathcal{E}^{\star}_i$ is the contribution to the self-energy that originates from the poles in Eq.~(\ref{eqn:ApproxPoles}) that arise for an excited state $i$ for which $\w_{mi}<0$. Expressions equivalent to Eqs.~(\ref{eqn:ShiftFinal}) and (\ref{eqn:ShiftFinal1}) have been derived before by different methods, e.g. by linear response theory \cite{McLachlan,sipe} or later by the noise-current approach to phenomenological QED \cite{Yoshi}. 

The shift $\Delta \mathcal{E}_i$ is real because it is a convolution of atom and field susceptibilities which are real at complex frequencies \cite{McLachlan}. However, $\Delta \mathcal{E}_i^{\star}$ is complex; its imaginary part modifies the decay rates of excited states. In summary, we have
\begin{eqnarray}
\Delta E_i&=&\Delta \mathcal{E}_i+{\rm Re}\left( \Delta \mathcal{E}_i^{\star}\right)\nonumber\\
\Delta \Gamma_i&=&-\frac{2}{\hbar}{\rm Im}\left( \Delta \mathcal{E}_i^{\star}\right)\label{eqn:RatesDef}
\end{eqnarray}
where $\Delta E_i$ are the renormalized energy-level shifts and $\Delta \Gamma_i$ are the changes in decay rates.

\section{Energy-level shifts near a half-space}\label{sec:HSShifts}
\subsection{Ground state}\label{sec:GrounState}
Substituting the photon propagator (\ref{eqn:PhotonPropDiag}) into Eq.~(\ref{eqn:ShiftFinal}) we find that the energy shift of the atomic ground state $|g\rangle$ is given by
\begin{eqnarray}
\Delta E_g=-\frac{1}{8\pi^2\epsilon_0}\sum_m\int_0^\infty\rd q_\|\: q_\|\int_{0}^\infty\rd\w\frac{\w_{mg}}{\w^2+\w_{mg}^2}
\frac{e^{-2\sqrt{q_\|^2+\w^2}\mathcal{Z}}}{\sqrt{q_\|^2+\w^2}}\nonumber\\
\times\left\{\left[(q_\|^2+\w^2)\bar{r}^\TM-\w^2\bar{r}^\TE\right]|\mu_{mg}^\parallel |^2+2q_\|^2\bar{r}^\TM|\mu_{mg}^\perp|^2\right\},\nonumber\\\label{eqn:HalfSpaceShiftFinal}
\end{eqnarray}
where we have used the notation $|\mu_{mi}^\parallel |^2=|\mu_{mi}^x |^2+|\mu_{mi}^y |^2$ and $|\mu_{mi}^\perp |^2=|\mu_{mi}^z |^2$. The reflection coefficients are as defined in Eq.~(\ref{eqn:Fresnel}). In terms of the new variables they read
\begin{eqnarray}
\bar{r}^\TE &=& \frac{\sqrt{\w^2+q_\|^2}-\sqrt{\epsilon(i\w)\w^2+q_\|^2}}{\sqrt{\w^2+q_\|^2}+\sqrt{\epsilon(i\w)\w^2+q_\|^2}},\nonumber\\
\bar{r}^\TM &=& \frac{\epsilon(i\w)\sqrt{\w^2+q_\|^2}-\sqrt{\epsilon(i\w)\w^2+q_\|^2}}{\epsilon(i\w)\sqrt{\w^2+q_\|^2}+\sqrt{\epsilon(i\w)\w^2+q_\|^2}}.\;\;\;
\label{eqn:BarFresnel}
\end{eqnarray}
Note that we have replaced $\xi(\w)$ by $\epsilon(\w)$ in Eq.~(\ref{eqn:BarFresnel}) compared to Eq.~(\ref{eqn:Fresnel}), because for the relevant frequencies both functions coincide (see Section \ref{sec:PhotonDress} and Appendix \ref{App:Epsilon}). If we now introduce polar coordinates according to, ${\w=\w_{mg}\:x\cos\phi, q_\|=\w_{mg}\:x\sin\phi}$ and then write $y=\cos\phi$, we obtain the perhaps most useful expression for the ground-state shift, especially for numerical analysis and for investigating the effects of retardation, 
\begin{eqnarray}
\Delta E_g=-\frac{1}{8\pi^2\epsilon_0}\sum_m\int_0^\infty\rd xx^3\int_{0}^1\rd y \frac{\w_{mg}^3}{1+x^2y^2}e^{-2\w_{mg}\mathcal{Z}x}\;\;\nonumber\\
\times\left[\left(\tilde{r}^\TM-y^2\tilde{r}^\TE\right)|\mu_{mg}^\parallel |^2+2\left(1-y^2\right)\tilde{r}^\TM|\mu_{mg}^\perp|^2\right].\hspace{.5 cm};\label{eqn:HalfSpaceShiftFinal2}
\end{eqnarray}
The result in Eq.~(\ref{eqn:HalfSpaceShiftFinal2}) formally takes the same form as the results obtained in calculations involving only non-dispersive dielectrics (see e.g.~\cite{Wu}), the only difference being the reflection coefficients that now, through the dielectric constant, depend on the product $xy$ of the integration variables which is the photon frequency in units of $\w_{mg}$,
\begin{eqnarray}
\tilde{r}^\TE&=&\frac{1-\sqrt{y^2[\epsilon(i\w_{mg}xy)-1]+1}}{1+\sqrt{y^2[\epsilon(i\w_{mg}xy)-1]+1}},\nonumber\\
\tilde{r}^\TM&=&\frac{\epsilon(i\w_{mg}xy)-\sqrt{y^2[\epsilon(i\w_{mg}xy)-1]+1}}{\epsilon(i\w_{mg}xy)+\sqrt{y^2[\epsilon(i\w_{mg}xy)-1]+1}}\label{eqn:TildedFresnel}.
\end{eqnarray}
Eq.~(\ref{eqn:HalfSpaceShiftFinal2}) is suitable for numerical analysis but does not give immediate insight into the dependence of the energy shift as a function of the distance from the surface. It is therefore instructive to consider some of its limiting cases.

As has been spelled out e.g.~in Ref.~\cite{Wu}, the dimensionless parameter that plays a decisive role in the characteristics of the Casimir-Polder interaction is given by the combination $2\w_{mg}\mathcal{Z}$ which is the ratio of two time-scales: \emph{(i)} the typical time $2\mathcal{Z}/c$ needed by a virtual photon to make a round trip between the atom and the surface, and \emph{(ii)} the typical time-scale $\w_{mg}^{-1}$ at which the atomic system evolves. While Eq.~(\ref{eqn:HalfSpaceShiftFinal2}) includes a sum over atomic states $|m\rangle$, in reality contributions to the shift will be dominated by the state which is connected to the ground state by the strongest dipole transition. We shall call the frequency $\w_{mg}$ that pertains to this strongest transition the ``typical transition frequency'' and it is this number that enters the retardation criterion parameter. Roughly speaking, if $2\w_{mg}\mathcal{Z}\ll 1$ we are in the so-called nonretarded regime when the time needed by the photon to travel between the dielectric and the atom is negligibly small compared to the typical atomic time-scale. Then the interaction can safely be approximated as instantaneous and our result should reduce to that calculated by Barton \cite{Barton}, who considers only the Coulomb interaction of an atom with surface polaritons. In the opposite case, $2\w_{mg}\mathcal{Z}\gg 1$, the interaction becomes retarded, i.e. by the time the photon has completed a round trip, the atomic state has changed significantly. In that case, for reasons that are not obvious but will become apparent later, the interaction depends only on static polarizabilities, i.e.~the polarizabilities evaluated at zero frequency. The diagonal polarizability of the spherically symmetric atom is then
\begin{equation}
\alpha^i_{\nu\nu}(0)=\left.\sum_j \frac{2\w_{ji}\left|
\langle j|\mu_\nu|i\rangle\right|^2}{\w_{ji}^2-\omega^2}\right|_{\w=0}=2\sum_j \frac{\left|
\langle j|\mu_\nu|i\rangle\right|^2}{\w_{ji}}\;, \label{eqn:AtomicPolarizability}
\end{equation}
and the susceptibility of the dielectric becomes
\begin{equation}
\frac{\epsilon(0)}{\epsilon_0}=\left.1+\frac{\w_{\rm P}^2}{\w_{\rm T}^2-\w^2-2i\gamma\w}\right|_{\w=0}=1+\frac{\w_{\rm P}^2}{\w_{\rm T}^2}\;,\label{eqn:Polarizability}
\end{equation}
as explained in Appendix \ref{App:Epsilon}.

\subsubsection{Nonretarded limit}\label{par:HSNonRetGround}
In order to take the non-retarded limit of the energy shift it is best to start from Eq.~(\ref{eqn:HalfSpaceShiftFinal}). After changing variables from $\w$ to $s$ with ${\;\w=(2\w_{mg}\mathcal{Z}q_\|)\: s}$ we take the limit $2\w_{mg}\mathcal{Z}\rightarrow 0$, which we may do because the line $s=\infty$ does not contribute to the integral, and  approximate
\begin{eqnarray}
q_\|^2+\w^2&\rightarrow&q_\|^2\left[1+(2\w_{mg}\mathcal{Z})^2s^2\right]\approx q_\|,\nonumber\\
q_\|^2+\epsilon\:\w^2&\rightarrow&q_\|^2\left[1+\epsilon\, (2\w_{mg}\mathcal{Z})^2s^2\right]\approx q_\|.\nonumber
\end{eqnarray}
This significantly simplifies Eq.~(\ref{eqn:HalfSpaceShiftFinal}). The {$q_\|$ integral} becomes elementary, and the final result reads
\begin{eqnarray}
\Delta E^{\rm nonret}_g\approx-\frac{1}{32\pi^2\epsilon_0\mathcal{Z}^3}\sum_m\int_0^\infty\rd \w\;\frac{\w_{mg}}{\w^2+\w_{mg}^2}\;\frac{\epsilon(i\w)-1}{\epsilon(i\w)+1}\nonumber\\
\times\left(|\mu_{mg}^\parallel|^2+2|\mu_{mg}^\perp|^2\right).\hspace{1 cm}\label{eqn:NonRet}
\end{eqnarray}
We observe the expected $\mathcal{Z}^{-3}$ behaviour of the energy level shift in the non-retarded or ``van der Waals'' regime. In order to see that the result in Eq.~(\ref{eqn:NonRet}) is equivalent to the slightly more awkward principal-value integral given in Eq.~(7.14) of Ref.~\cite{Barton} or Eq.~(13) of Ref.~\cite{BartonComments}, one needs to re-write
$$
\frac{\w_{mg}}{\w^2+\w_{mg}^2} = \frac12 \left(\frac1{\w_{mg}-i\w}+\frac1{\w_{mg}+i\w} \right)
$$ 
and the re-rotate the contour from $\w$ to $i\w$ in the first and to $-i\w$ in the second summand. Eq.~ (\ref{eqn:NonRet}) also confirms the results derived on the basis of the phenomenological noise-current approach to quantum electrodynamics with dielectric media, see e.g.~\cite{Yoshi}.

\subsubsection{Retarded limit}
In order to work out an approximation to the energy shift when retardation is dominant, it is convenient to start with Eq.~(\ref{eqn:HalfSpaceShiftFinal2}) where the decisive parameter $2\w_{mg}\mathcal{Z}$ is present in the exponential which in the limit $2\w_{mg}\mathcal{Z}\rightarrow\infty$ strongly damps the integrand. Then the main contributions to the integral come from the neighbourhood of $x=0^+$ and one can obtain an asymptotic expansion of the integral by expanding the integrand in a Taylor series around this point. A straightforward calculation gives 
\begin{equation}
\Delta E^{\rm ret}_g\approx-\frac{3}{64\pi^2\epsilon_0}\sum_m\sum_{\sigma=\parallel,\perp}
\left(\frac{c^\sigma_4}{\mathcal{Z}^4}-\frac{4\gamma}{\w_{\rm T}^2}\frac{c^\sigma_5}{\mathcal{Z}^5}\right)
\frac{|\langle g|\mu^\sigma|m\rangle|^2}{\w_{mg}},\label{eqn:retardedHS}
\end{equation}
where in the parentheses we have neglected terms of order $\w_{mg}^{-2}\mathcal{Z}^{-6}$ and higher. The coefficients  $c^\sigma_{4,5}$, to be given below, depend only on the static dielectric constant of the material. The fact that to leading order the Casimir-Polder force depends only on the static polarizability of the atom, Eq.~(\ref{eqn:AtomicPolarizability}), is well known \cite{McLachlan}. Therefore the leading-order $\mathcal{Z}^{-4}$ term in Eq.~(\ref{eqn:retardedHS}) is identical to the retarded limit of the energy shift of a ground-state atom interacting with a non-absorptive dielectric half-space described by a static refractive index ${n(0)\equiv n =\sqrt{1+\w_{\rm P}^2/\w_{\rm T}^2}}$, which has been derived previously \cite{Wu}. We just quote the results for the coefficients $c_4^{\parallel,\perp}$ from Ref.~\cite{Wu}:
\begin{eqnarray}
c_4^\parallel=-\frac{1}{n^2-1}\left(\frac{2}{3}n^2+n-\frac{8}{3}\right)\hspace{2 cm}\nonumber\\
+\frac{2n^4}{(n^2-1)\sqrt{n^2+1}}\ln\left(\frac{\sqrt{n^2+1}+1}{n\left[\sqrt{n^2+1}+n\right]}\right)\nonumber\\
\;\;+\frac{2n^4-2n^2-1}{(n^2-1)^{3/2}}\ln\left(\sqrt{n^2+1}+n\right),\nonumber\\
c_4^\perp=\frac{1}{n^2-1}\left(4n^4-2n^3-\frac{4}{3}n^2+\frac{4}{3}\right)\hspace{1 cm}\nonumber\\
-\frac{4n^6}{(n^2-1)\sqrt{n^2+1}}\ln\left(\frac{\sqrt{n^2+1}+1}{n\left[\sqrt{n^2+1}+n\right]}\right)\nonumber\\
\;\;-\frac{2n^2(2n^4-2n^2+1)}{(n^2-1)^{3/2}}\ln\left(\sqrt{n^2-1}+n\right).\nonumber
\end{eqnarray}
In other words, to leading-order, in the retarded limit, absorption makes no difference and only static polarizabilities, of both the dielectric and the atom, matter. This is because the photon wavelengths that matter the most in the atom-wall interaction are of the order of the distance between the atom and the surface of the dielectric and longer. Thus for an atom in the so-called far-zone only long wavelengths of the electromagnetic radiation come into play, which means low frequencies.

\begin{figure}[htbp]
  \centering
    \hspace{-.5 cm}\includegraphics[width=9.0 cm, height=7.7 cm]{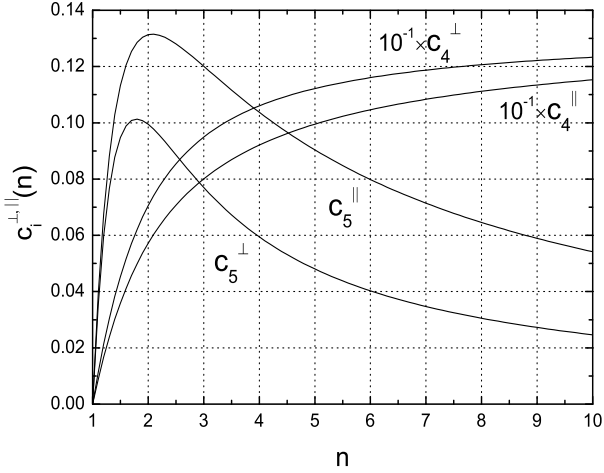}
  \caption{Plot of the coefficients $c_i^{\sigma}(n)$ that enter Eq.~(\ref{eqn:retardedHS}) for different values of the static refractive index $n\equiv n(0)$.}
  \label{fig:FFunsHS}
\end{figure}

\begin{figure}[htbp]
  \centering
    \hspace{.9 cm}\includegraphics[width=9.0 cm, height=7.7 cm]{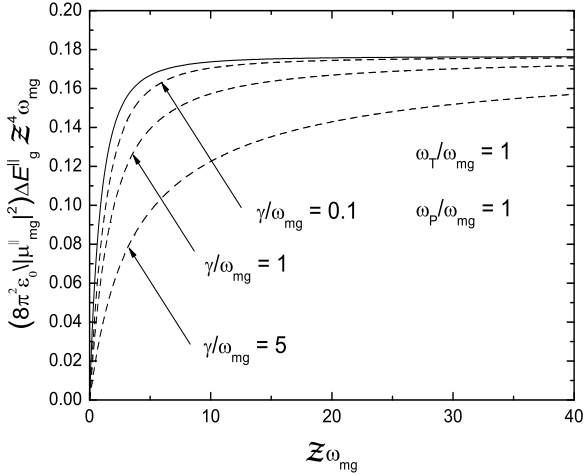}
  \caption{Plot of the exact ground-state energy shift (contributions due to the perpendicular component of the atomic dipole moment) $\Delta E^\parallel_g$, Eq.~(\ref{eqn:HalfSpaceShiftFinal}), multiplied by $\mathcal{Z}^4\w_{mg}$ as a function of $\mathcal{Z}\w_{mg}$ for various values of the damping parameter $\gamma$. Solid line represents the energy shift caused by the non-absorptive and non-dispersive dielectric half-space with static refractive index $n(0)=\sqrt{2}$.}
  \label{fig:HalfSpaceShift}
\end{figure}

Now we turn our attention to the next term in the asymptotic expansion which is proportional to $\mathcal{Z}^{-5}$. This is the first term that contains information about corrections to the energy shift due to absorption in the retarded regime. Apart from the factor $4\gamma/ \w_{\rm T}^2$, the dimensionless coefficients $c_5^{\sigma}$ depend again only on the static refractive index $n=\sqrt{1+\w_{\rm P}^2/\w_{\rm T}^2}$ and are given by
\begin{eqnarray}
c_5^\parallel=\frac{1}{3(n-1)(n+1)^2(n^2+1)}\hspace{4 cm}
\nonumber\\
\times\bigg\{6n^6-3n^5-11n^4+4n^3+2n^2-5n+7\hspace{1.5 cm}\nonumber\\
-6n^2\big(n^5+n^4-n^3-n^2-2n-2\big)\ln\left[n\left(\frac{n+1}{n^2+1}\right)\right]\bigg\},\nonumber\\
c_5^\perp=\frac{4}{3(n-1)(n+1)^2(n^2+1)}\hspace{4 cm}
\nonumber\\
\times\bigg\{-6n^8+3n^7+10n^6-5n^5+3n^4-n^3-6n^2+n+1\nonumber\\
+3n^4\big(2n^5+2n^4-n^3-n^2-3n-3\big)\ln\left[n\left(\frac{n+1}{n^2+1}\right)\right]\bigg\}.\nonumber\\
\label{eqn:HSCoeffs}
\end{eqnarray}
We provide plots of these functions in Fig.~\ref{fig:FFunsHS} from where a quick estimate of the value of these coefficients can be obtained.  Since both $c_5^\parallel$ and $c_5^\perp$ are positive we see that absorption reduces the magnitude of the ground-state energy shift by an amount that is proportional to the damping constant $\gamma$, cf. Fig.~\ref{fig:HalfSpaceShift}. We also note that the correction goes with the inverse square of the absorption frequency $\w_{\rm T}$ in the dielectric so that only absorption lines that lie at sufficiently low frequencies make a significant difference. This happens because the main contribution to the shift of the ground state in the retarded limit comes from long wavelengths or equivalently small values of $x$ (which is a scaled frequency). Therefore, the integral is not sensitive to any absorption peaks which lie at higher frequencies as there the integrand is highly damped anyway, cf. Eq. (\ref{eqn:HalfSpaceShiftFinal2}).

\subsection{Excited states.}\label{sec:HSEcxited}
The shift of an excited energy level gets contributions from both parts of $\Delta \mathcal{E}^{\rm ren}$, Eqs.~(\ref{eqn:ShiftFinal}) and (\ref{eqn:ShiftFinal1}). The non-residue contributions, Eq.~(\ref{eqn:ShiftFinal}), assume exactly the same form as the results of the previous section. Therefore we will not analyze them again but shall instead have a closer look at the additional contributions due to Eq.~(\ref{eqn:ShiftFinal1}). 

Plugging in the photon propagator, Eq.~(\ref{eqn:PhotonPropDiag}), we find that the energy shift of the excited state $|i\rangle$ is given by the real part of the following expression
\begin{eqnarray}
\Delta \mathcal{E}_i^{\star}=-\frac{i}{8\pi\epsilon_0}\sum_{m<i}\int_0^\infty\rd q_\| \frac{q_\| e^{2i\mathcal{Z}\sqrt{\w_{mi}^2-q_\|^2}}}{\sqrt{\w_{mi}^2-q_\|^2+i\eta}}\hspace{2.1 cm}\nonumber\\
\times
\left\{ \left[\w_{mi}^2r^\TE_{mi}-(\w_{mi}^2-q_\|^2)r^\TM_{mi}\right]|\mu_{mi}^\parallel|^2+2q_\|^2r^\TM_{mi}|\mu_{mi}^\perp|^2\right\}.\nonumber\\ \label{eqn:ResidueContributions}
\end{eqnarray}
Here $r^\lambda_{mi}$ are the reflection coefficients of Eq.~(\ref{eqn:Fresnel}) evaluated at the atomic transition frequencies ${\w=|\w_{mi}}|$. Also, the restriction of the sum over atomic states to those lying below the state $|i\rangle$ should be noted. For the purposes of asymptotic analysis of $\Delta \mathcal{E}_i^{\star}$ we change the integration variable in Eq.~(\ref{eqn:ResidueContributions}) to ${k_z=\sqrt{\w_{mi}^2-q_\|^2}}/|\w_{mi}|$ and get
\begin{eqnarray}
\Delta \mathcal{E}_i^{\star}=\frac{i}{8\pi\epsilon_0}\sum_{m<i}|\w_{mi}|^3\int_1^{i\infty}\rd k_z e^{2i|\w_{mi}|\mathcal{Z}k_z}\;\hspace{1.9 cm}\nonumber\\
\times\left\{ \left(\bar{r}^\TE_{mi}-k_z^2\bar{r}^\TM_{mi}\right)|\mu_{mi}^\parallel|^2+2\left(1-k_z^2\right)\bar{r}^\TM_{mi}|\mu_{mi}^\perp|^2\right\},\hspace{.7 cm}
\label{eqn:ResidueContributions1}
\end{eqnarray}
where the contour of integration runs from $k_z=1$ along the real axis to $k_z=0$ and then up along the imaginary axis to $k_z=i\infty$. The reflection coefficients expressed as functions of $k_z$ are
\begin{eqnarray}
\bar{r}^\TE_{mi}(k_z)&=&\frac{k_z-\sqrt{[\epsilon(|\w_{mi}|)-1]+k_z^2}}{k_z+\sqrt{[\epsilon(|\w_{mi}|)-1]+k_z^2}},\nonumber\\
\bar{r}^\TM_{mi}(k_z)&=&\frac{\epsilon(|\w_{mi}|)k_z-\sqrt{[\epsilon(|\w_{mi}|)-1]+k_z^2}}{\epsilon(|\w_{mi}|)k_z+\sqrt{[\epsilon(|\w_{mi}|)-1]+k_z^2}}.\label{eqn:ChangedFresnels}
\end{eqnarray}
We now go on to analyse $\Delta \mathcal{E}_i^{\star}$ in the nonretarded and retarded limits.

\subsubsection{Nonretarded limit} 
In the nonretarded limit of Eq.~(\ref{eqn:ResidueContributions1}) we have $2|\w_{mg}|\mathcal{Z}\ll 1$. It is expedient to split the integration in Eq. (\ref{eqn:ResidueContributions1}) in the following way
\begin{equation}
\int_1^{i\infty}\rd k_z=\int_0^{\infty}\rd (ik_z)-\int_0^{1}\rd k_z\label{eqn:Split}
\end{equation}
and note that in the limit $2|\w_{mi}|\mathcal{Z}\rightarrow 0$ the second integral on the RHS contributes to the asymptotic series only terms that are proportional to non-negative powers of $\mathcal{Z}$ and can therefore be discarded. The remaining part is given by
\begin{eqnarray}
\Delta \mathcal{E}_i^{\star,1}=-\frac{1}{8\pi\epsilon_0}\sum_{m<i}|\w_{mi}|^3\int_0^\infty \rd k_z e^{-2|\w_{mi}|\mathcal{Z}k_z}\hspace{1.7 cm}\nonumber\\
\times\left[(\tilde{r}^\TE_{mi}+k_z^2\tilde{r}^\TM_{mi})|\mu_{mi}^\parallel|^2+2(1+k_z^2)\tilde{r}^\TM_{mi}|\mu_{mi}^\perp|^2\right]\hspace{1 cm}\label{eqn:AlongRealAxis}
\end{eqnarray}
where $\tilde{r}^\lambda_{im}$ are the reflection coefficients of Eq.~(\ref{eqn:ChangedFresnels}) evaluated at imaginary argument $\tilde{r}^\lambda_{mi}=\bar{r}^\lambda_{mi}(ik_z)$. Scaling the integration variable according to $x=2|\w_{mi}|\mathcal{Z}\;k_z$ and approximating
\begin{equation}
\sqrt{[\epsilon(|\w_{mi}|)-1]-\frac{x^2}{(2|\w_{mi}|\mathcal{Z})^2}}\approx\frac{ix}{2|\w_{mi}|\mathcal{Z}}\;,
\end{equation}
we derive that, in the nonretarded limit, Eq.~(\ref{eqn:ResidueContributions}) becomes
\begin{eqnarray}
\Delta \mathcal{E}_i^{\star,{\rm nonret}}=-\frac{1}{32\pi\epsilon_0\mathcal{Z}^3}\sum_{m<i}\frac{\epsilon(|\w_{mi}|)-1}{\epsilon(|\w_{mi}|)+1}\hspace{1 cm}\nonumber\\
\left(|\mu_{mi}^\parallel|^2+2|\mu_{mi}^\perp|^2\right).\;\;\;\label{eqn:ExcitedNonRet}
\end{eqnarray}
To leading order the residue contributions to the energy shift of the excited state $|i\rangle$, cf. Eq. (\ref{eqn:ShiftFinal1}), are given by the real part of the above expression,
\begin{eqnarray}
\Delta E_i^{\star, {\rm nonret}}=-\frac{1}{32\pi\epsilon_0\mathcal{Z}^3}\sum_{m<i}\frac{|\epsilon(|\w_{mi}|)|^2-1}{|\epsilon(|\w_{mi}|)+1|^2}\hspace{1 cm}\nonumber\\
\left(|\mu_{mi}^\parallel|^2+2|\mu_{mi}^\perp|^2\right).\;\;\;\label{eqn:ReExcitedNonRet}
\end{eqnarray}
Thus in the nonretarded regime the residue contributions behave as $\mathcal{Z}^{-3}$ and therefore are of the same order as the non-residue contributions, cf. Eq. (\ref{eqn:NonRet}). The result in Eq.~(\ref{eqn:ReExcitedNonRet}) is in fact equivalent to the real part of Eq.~(7.10) derived in Ref.~\cite{Barton}.

\subsubsection{Retarded limit} Now we turn our attention to the asymptotic behaviour of Eq.~(\ref{eqn:ResidueContributions1}) in the retarded limit, i.e.~when $2|\w_{mi}|\mathcal{Z} \gg 1$. It is again useful to split the integration in the same way as in Eq.~(\ref{eqn:Split}), only that now both integrals play an important role. The first contribution, the integral along $k_z \in [0,i\infty]$, given in Eq.~(\ref{eqn:AlongRealAxis}), can be tackled by use of Watson's lemma \cite{Bender}. Noting that for $2|\w_{mi}|\mathcal{Z} \gg 1$ the integrand is strongly damped, we separate off the exponential and expand the remaining part into Taylor series about $k_z=0$. The resulting integrals are elementary and we obtain for the leading term
\begin{eqnarray}
\Delta \mathcal{E}_i^{\rm \star,1,ret}=\frac{1}{8\pi\epsilon_0}\sum_{m<i}|\w_{mi}|^3\bigg\{\frac{|\mu_{mi}^\parallel|^2}{2|\w_{mi}|\mathcal{Z}}\hspace{2 cm}\nonumber\\
+2\left[1-\frac{2i\epsilon(|\w_{mi}|)}{\sqrt{\epsilon(|\w_{mi}|)-1}}\frac{1}{2|\w_{mi}|\mathcal{Z} }\right]\frac{|\mu_{mi}^\perp|^2}{2|\w_{mi}|\mathcal{Z}}\bigg\}.\;\;\;\;\;\label{eqn:contributions1}
\end{eqnarray}
Next we deal with the integral on the interval $k_z \in [0,1]$ which, unlike in the nonretarded case, cannot be discarded. However, its asymptotic expansion in inverse powers of $\mathcal{Z}$ is easily obtained by repeated integration by parts. Interestingly, the asymptotic series contain non-oscillatory terms that exactly cancel out the contributions given in Eq.~(\ref{eqn:contributions1}). Altogether we find that the leading-order and next-to-leading-order terms are
\begin{eqnarray}
\Delta \mathcal{E}_i^{\star \rm,ret}=\frac{1}{4\pi\epsilon_0} \sum_{m<i}|\w_{mi}|^3\frac{n(|\w_{mi}|)-1}{n(|\w_{mi}|)+1}\;e^{2i|\w_{mi}|\mathcal{Z}}\nonumber\\
\times\left[\frac{|\mu_{mi}^\parallel|^2}{2|\w_{mi}|\mathcal{Z}} +2i\frac{|\mu_{mi}^\perp|^2}{(2|\w_{mi}|\mathcal{Z})^2}\right],\;\;\;\label{eqn:ExcitedRet}
\end{eqnarray}
with the refractive index $n(|\w_{mi}|)\equiv\sqrt{\epsilon(|\w_{mi}|)}$. 
It is interesting to observe that to leading-order in $\mathcal{Z}$ only contributions due to the parallel component of the atomic dipole moment are contributing; contributions due to the perpendicular component of the atomic dipole moment appear only in next-to-leading order. Again we need to take the real part of Eq.~(\ref{eqn:ExcitedRet}) to get the explicit form of the energy shift 
\begin{eqnarray}
\Delta E_i^{\rm \star,ret}=\frac{1}{4\pi\epsilon_0}\sum_{m<i}\frac{|\w_{mi}|^3}{|n(\w_{mi})+1|^2}\hspace{3 cm}
\nonumber\\
\times\bigg\{\bigg[(|n(|\w_{mi}|)|^2-1)\cos(2|\w_{mi}|\mathcal{Z})\hspace{2.8 cm}
\nonumber\\
-2{\rm Im}[n(|\w_{mi}|)]\sin(2|\w_{mi}|\mathcal{Z})\bigg]\frac{|\mu_{mi}^\parallel|^2}{2|\w_{mi}|\mathcal{Z}}\hspace{1.5 cm}
\nonumber\\
-2\bigg[(|n(|\w_{mi}|)|^2-1)\sin(2|\w_{mi}|\mathcal{Z})\hspace{2.8 cm}
\nonumber\\
+2{\rm Im}[n(|\w_{mi}|)]\cos(2|\w_{mi}|\mathcal{Z})\bigg]\frac{|\mu_{mi}^\perp|^2}{(2|\w_{mi}|\mathcal{Z})^2}\bigg\}.\hspace{.7 cm}\label{eqn:ExitedFinalRet}
\end{eqnarray}
We see that in the retarded regime the two contributions to the shift of an excited state behave quite differently. The non-residue contribution in Eq.~(\ref{eqn:ShiftFinal}) behaves as $\mathcal{Z}^{-4}$ (see the analysis of the ground state shift in Section \ref{sec:GrounState}), and the residue contribution in Eq.~(\ref{eqn:ExitedFinalRet}) depends on distance as $\mathcal{Z}^{-1}$. While it would be tempting to jump to the conclusion that Eq.~(\ref{eqn:ExitedFinalRet}) will always dominate, this might in fact not always be the case as the relative size of the two contributions also depends on the values of the dipole matrix elements involved, which can vary significantly. Furthermore, Eq.~(\ref{eqn:ExitedFinalRet}) is oscillatory, so that at least in principle there are sets of parameters for which it vanishes. Finally, we remark that it is easy to verify that in the limit of non-absorptive dielectric media our results reduce to those derived in Ref.~\cite{Wu}. 

\section{Spontaneous decay rates near a half-space}
\begin{figure}[htbp]
  \centering
  \begin{tabular}{c}
\includegraphics[width=7 cm, height=4.5 cm]{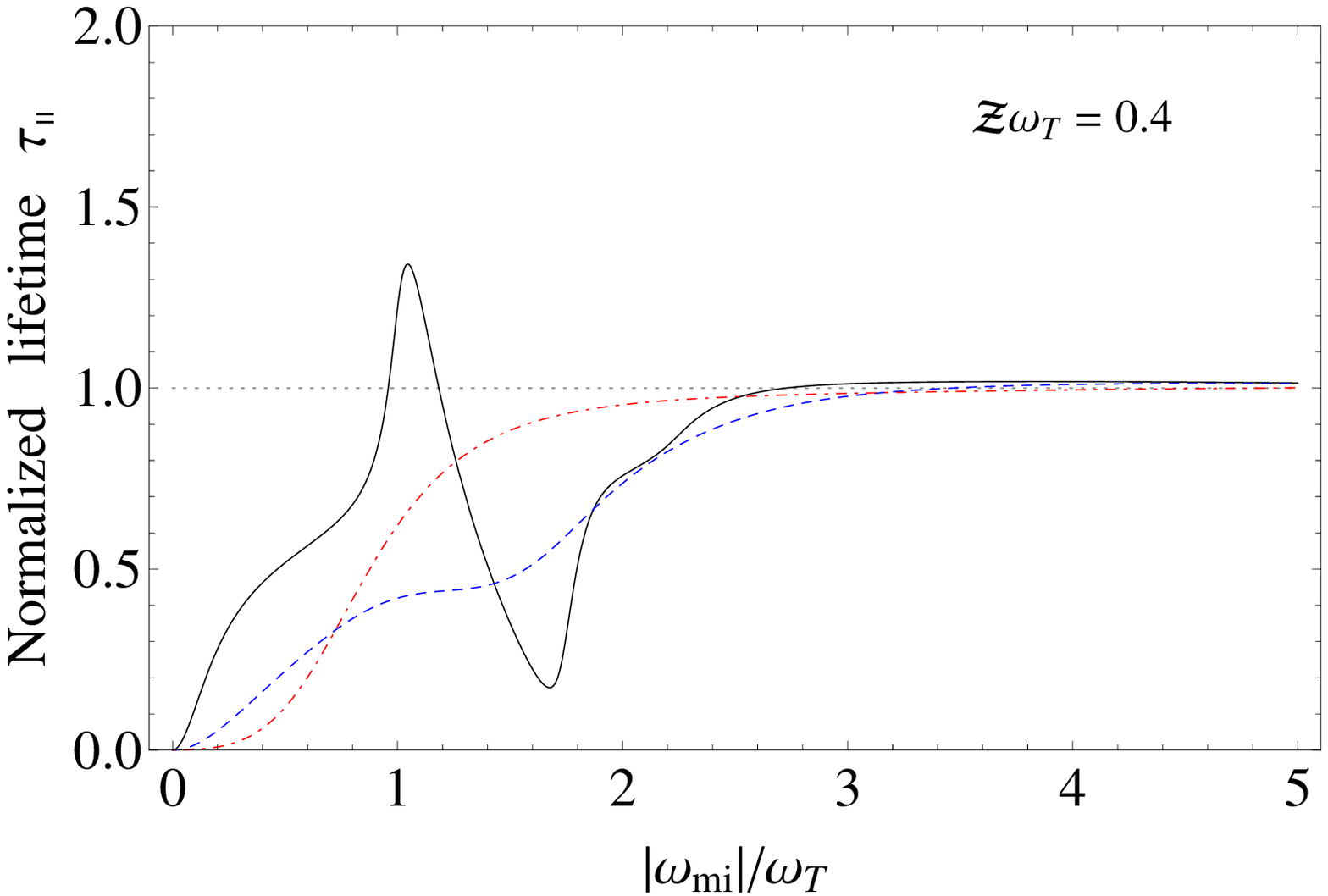} \\
\includegraphics[width=7 cm, height=4.5 cm]{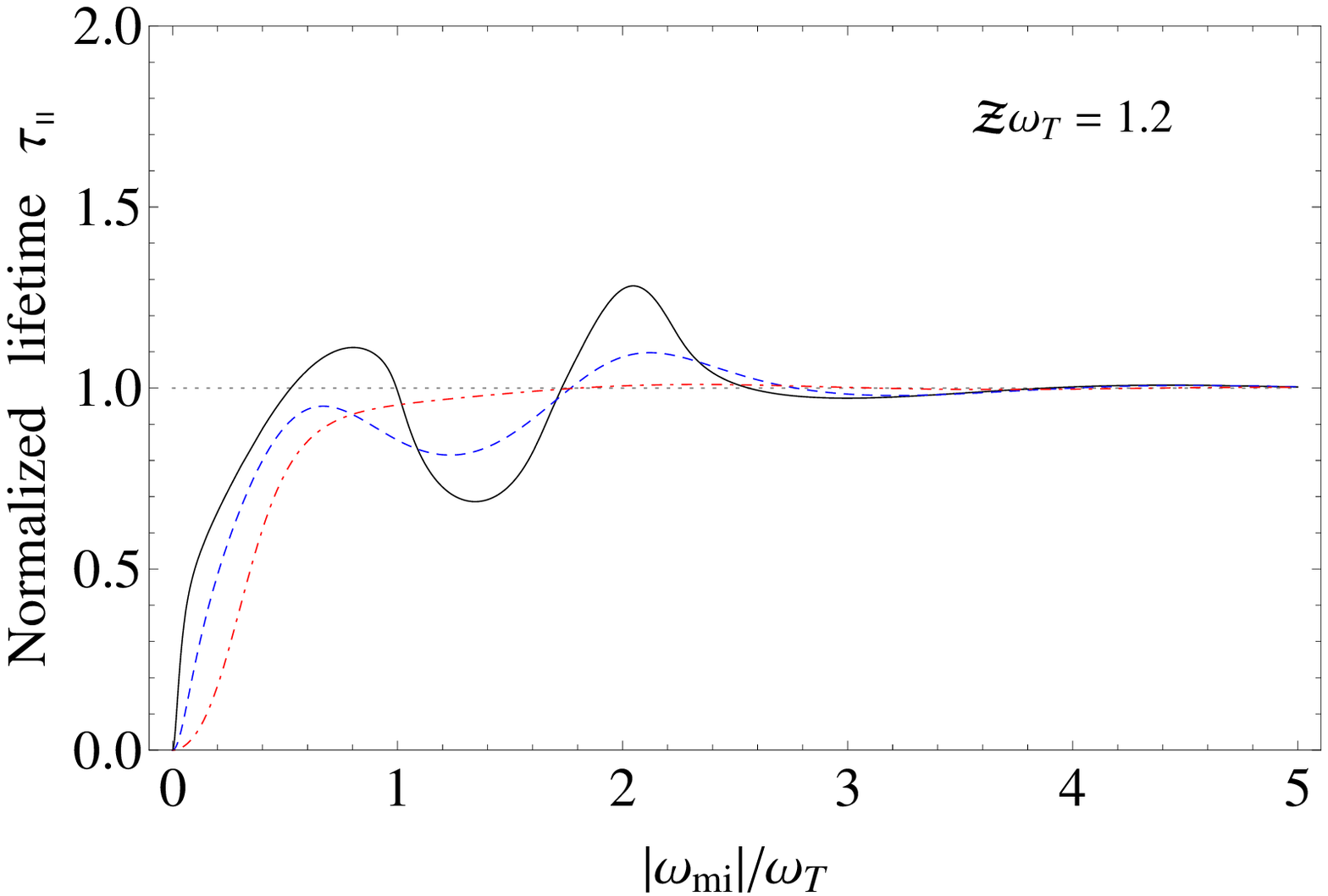} \\
\includegraphics[width=7 cm, height=4.5 cm]{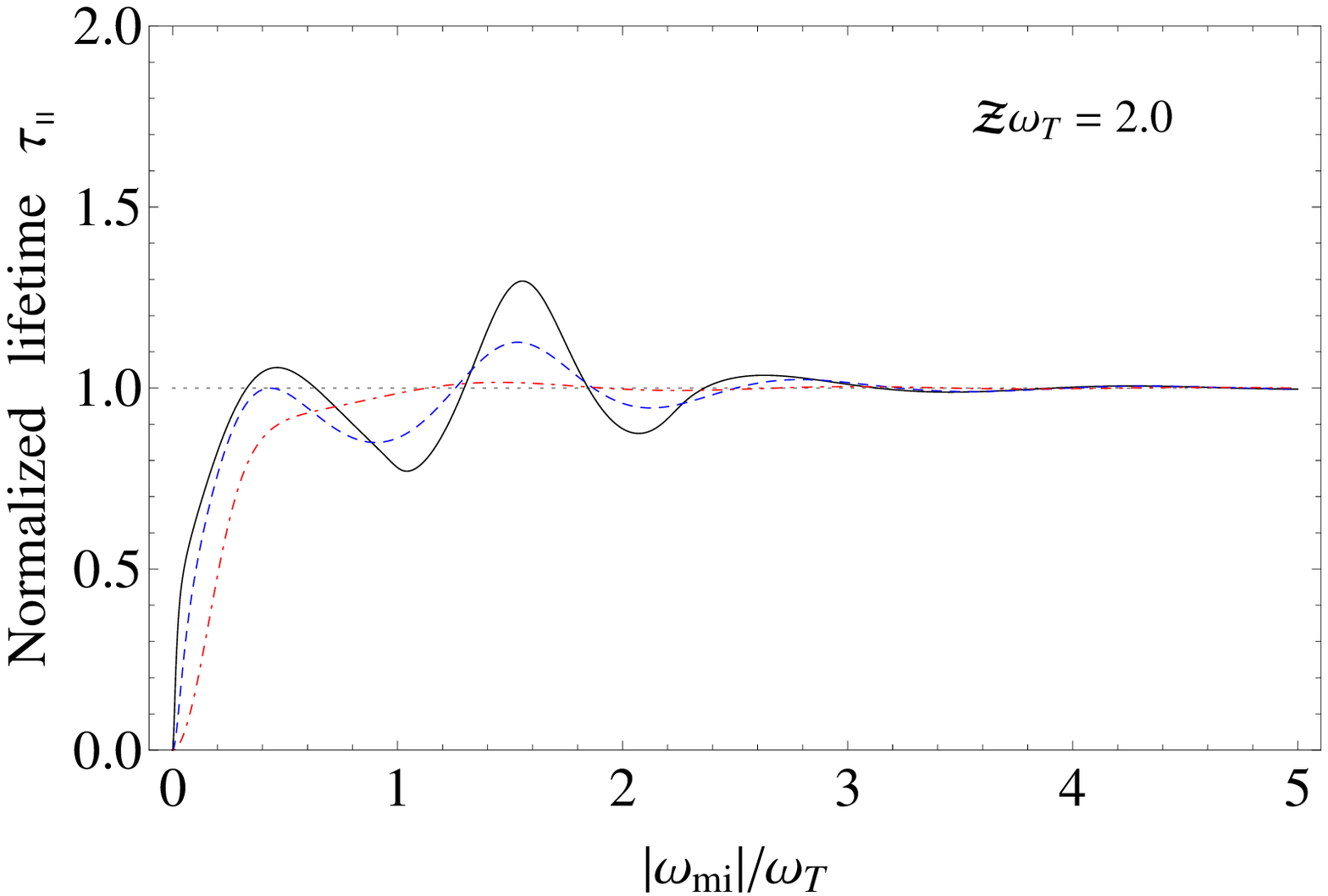} \\
\includegraphics[width=7 cm, height=4.5 cm]{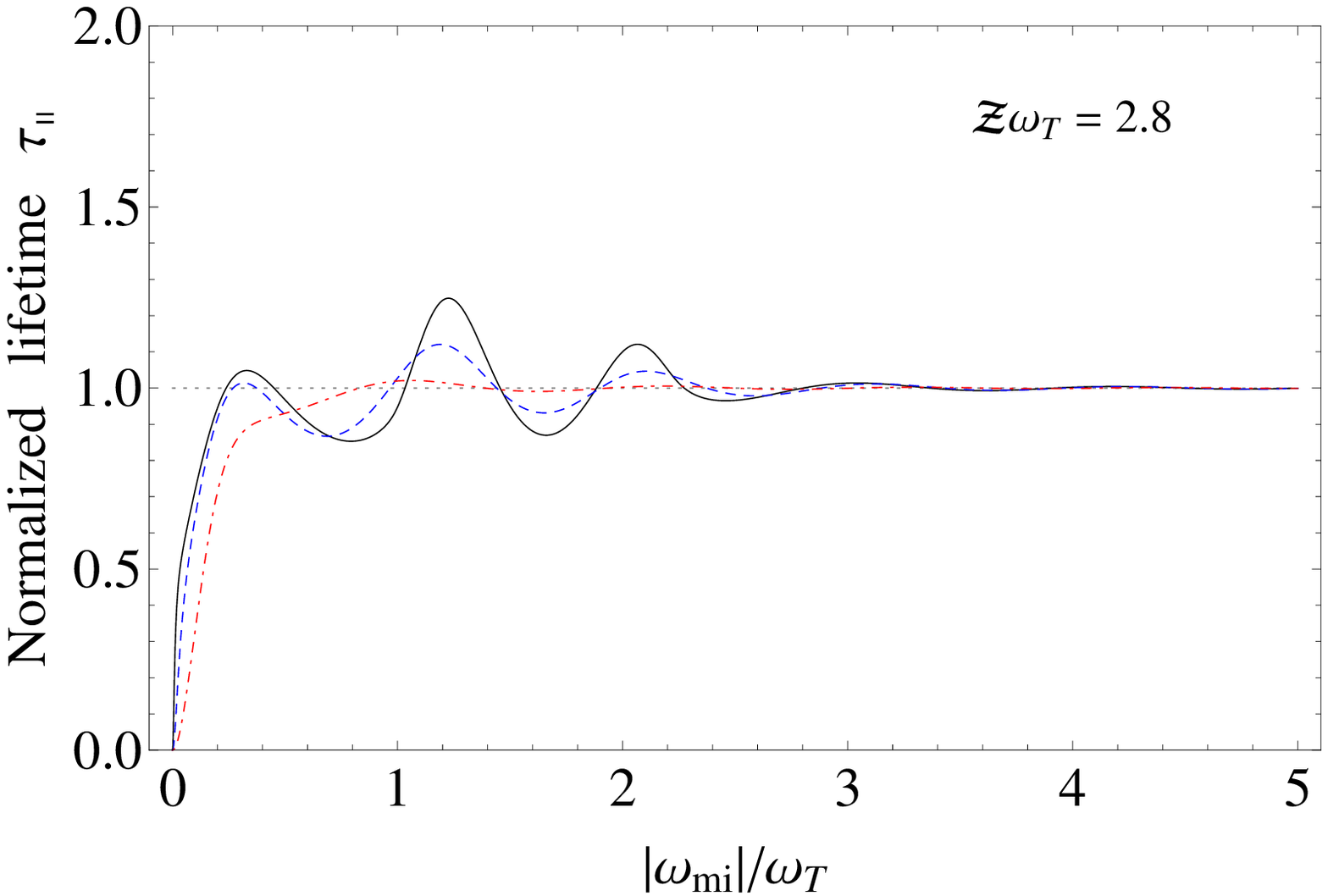} \\
\end{tabular}
\caption[Decay rates.]{\label{fig:HalfSpaceLifetime}{Normalized lifetime $\tau_{\parallel}$, Eq. (\ref{eqn:NormalLife}), of the atomic state $| i \rangle$ plotted as a function of $|\w_{mi}|/\w_{\rm T}$. The sequence of graphs corresponds to various distances of the atom from the mirror $\mathcal{Z}\w_{\rm T}$, as indicated. The three different line styles (colours) indicate distinct choices of the damping constant $\gamma$ in the dielectric: $\gamma/\w_{\rm T}=0.05$ (black, solid), $\gamma/\w_{\rm T}=0.5$ (blue, dashed), $\gamma/\w_{\rm T}=5$ (red, dot-dashed). For sufficiently high frequencies $|\w_{mi}|$ the dielectric becomes transparent. If the atom is close to the surface and absorption is small, the interaction is resonant at $|\w_{mi}|/\w_{\rm T}\approx 1$, i.e.~when the frequency of the atomic transition coincides with the absorption line of the dielectric.}  }
 \label{fig:HalfSpace}
\end{figure}

The spontaneous decay rates are given by the imaginary part of the complex self-energy, Eq.~(\ref{eqn:RatesDef}).
As the non-residue contributions to the self-energy in Eq.~(\ref{eqn:ShiftFinal}) are real, these contribute towards the energy-level shifts only, and the decay rates are contained solely in the residue contributions to the self-energy, Eq.~(\ref{eqn:ResidueContributions}), which are complex. In the non-retarded limit the decay rates are given by the imaginary part of Eq.~(\ref{eqn:ExcitedNonRet}),
\begin{equation}
\Delta\Gamma^{\rm nonret}_i=\frac{1}{8\pi\epsilon_0\mathcal{Z}^3}\sum_{m<i}\frac{{\rm Im}[\epsilon(|\w_{mi}|)]}{|\epsilon(|\w_{mi}|)+1|^2}\left(|\mu_{mi}^\parallel|^2+2|\mu_{mi}^\perp|^2\right),\label{eqn:HSDecayNonRet}
\end{equation}
and in the retarded limit by the imaginary part of Eq.~(\ref{eqn:ExcitedRet}):
\begin{eqnarray}
\Delta \Gamma^{\rm ret}_i=-\frac{1}{2\pi\epsilon_0}\sum_{m<i}\frac{|\w_{mi}|^3}{|n(\w_{mi})+1|^2}\hspace{2.8 cm}\nonumber\\
\times\bigg\{\bigg[(|n(|\w_{mi}|)|^2-1)\sin(2|\w_{mi}|\mathcal{Z})\hspace{2.5 cm}\nonumber\\
+2{\rm Im}[n(|\w_{mi}|)]\cos(2|\w_{mi}|\mathcal{Z})\bigg]\frac{|\mu_{mi}^\parallel|^2}{2|\w_{mi}|\mathcal{Z}}\hspace{0.8 cm}\nonumber\\
+2\bigg[(|n(|\w_{mi}|)|^2-1)\cos(2|\w_{mi}|\mathcal{Z})\hspace{2.5 cm}\nonumber\\
-2{\rm Im}[n(|\w_{mi}|)]\sin(2|\w_{mi}|\mathcal{Z})\bigg]\frac{|\mu_{mi}^\perp|^2}{(2|\w_{mi}|\mathcal{Z})^2}\bigg\}.\hspace{0.6 cm}\label{eqn:HSDecayRet}
\end{eqnarray}
The result in Eq.~(\ref{eqn:HSDecayNonRet}) is found to be in an agreement with that derived in Ref.~\cite{Yeung}, their Eq. (128). A consistency check on Eq.~(\ref{eqn:HSDecayRet}) is that it reduces to the results given in Ref.~\cite{Wu} if we assume $n(\w)$ to be real and frequency-independent. 

As a numerical example we plot the normalized lifetime of the atomic state $|i\rangle$ decaying into a lower state $|m\rangle$. For simplicity we assume a two-level system and $|\mu_{mi}^\perp|=0$ so that the atom is polarized horizontally with respect to the surface. Then the normalized lifetime that we plot in Fig.~\ref{fig:HalfSpaceLifetime} is given by 
\begin{equation}
\tau_\parallel^{-1}=\frac{\Delta\Gamma_i}{\Delta\Gamma_i^0}=-\dfrac{2\;{\rm Im}\left(\Delta \mathcal{E}^\star_i\right)}{\hbar\Delta\Gamma_i^{0}}\label{eqn:NormalLife}
\end{equation}
where the quantity $\Delta\mathcal{E}^\star_i$ comes from Eq.~(\ref{eqn:ResidueContributions}) and $\Delta\Gamma_i^{0}$ is the well-known decay rate in free-space
\begin{equation}
\Delta\Gamma_i^{0}=\frac{|\w_{mi}|^3|\mu_{mi}|^2}{3\pi\epsilon_0\hbar}.
\end{equation}

\section{Summary and conclusions}
We have shown that starting from a gauge-independent microscopic model as represented by the Hamiltonian (\ref{eqn:TotalH}) it is possible to develop a formalism which allows to calculate QED corrections in the presence of absorptive and dispersive boundaries. We have used a diagrammatic technique to integrate out the damped polaritons in order to arrive at a Dyson equation for the electromagnetic displacement field propagator. We have solved this integral equation exactly, using traceable methods. The knowledge of the exact propagator has enabled us to calculate analytically the one-loop self-energy diagram for an electron bound in an atom near a dielectric half-space and hence to determine its energy-level shifts and the change in transition rates, which derive from the real and imaginary part of the electron's self-energy, respectively. This is serving as a proof of principle that the theoretical framework developed here works correctly and efficiently, as most of these results have, in one form or another, been derived previously by other methods, though often with considerably more effort or much less rigour, especially as regards basic principles.

We have looked at the role of the material's absorption in some detail and confirmed the previously known result that absorption has the most profound impact on the atomic system in the non-retarded regime, i.e.~when the distance between the atom and the mirror is much smaller than the wavelength of the dominant atomic dipole transition. If the distance between the atom and the surface far exceeds the wavelength of this dominant transition, then, to leading order, dispersion and absorption do not affect the \emph{ground-state} shift for which only static polarizabilities matter. The next-to-leading order corrections are proportional to damping constant of the Lorentz-type dielectric function, and it turns out that only the material's absorption lines that lie in the low-frequency region have a significant impact on the ground-state energy-level shift. We have also rederived the distance dependence of the excited energy-level shifts and spontaneous decay rates. We have confirmed the fact that in the non-retarded regime the absorption is of fundamental importance to both the change in decay rates and the energy-level shifts. For example, for an atom near a nondispersive dielectric the spontaneous decay rate in the near-zone comes out as a distance-independent constant \cite{Wu}, whereas in reality, when the absorption is taken into account, a distance dependence $\propto\mathcal{Z}^{-3}$ is obtained. In the far-zone or retarded limit, the presence of absorption does not affect the characteristic $\mathcal{Z}^{-1}$ behaviour of the excited energy-level shift and spontaneous decay rates, even though the coefficients differ from the nondispersive case.

\begin{acknowledgments}
It is a pleasure to thank Gabriel Barton for discussions. 
We would like to acknowledge financial support from the UK Engineering and
Physical Sciences Research Council.
\end{acknowledgments}

\appendix
\section{Dressed photon propagator as a boundary-value problem}
\label{App:DressProp}
We aim to show that the integral equation satisfied by the dressed photon propagator derived in Section \ref{sec:PhotonDress},
\begin{eqnarray}
D_{ik}(\br,\br';\w)=D^{(0)}_{ik}(\br-\br';\w)\hspace{3 cm}\nonumber\\
+\frac{K(\w)}{\epsilon^2_0}\int\rd^3\br_1g(\br_1)D^{(0)}_{ij}(\br-\br_1;\w)D_{jk}(\br_1,\br';\w).\;\;\label{eqn:IntegralEq}
\end{eqnarray}
can also be solved by considering it as a boundary-value problem.
Recall that $D^{(0)}_{ik}(\br-\br')$ is the photon propagator in free space, Eq.~(\ref{eqn:FreeSpacePosRep}), and $g(\br)$ is a dimensionless coupling constant that is equal to unity in the region occupied by the dielectric and vanishes otherwise. To describe a dielectric half-space occupying the $z<0$ region of space, as illustrated in Fig.~\ref{fig:hspace}, we take $g(\br_1)=\theta(-z_1)$ where $\theta$ is the Heaviside step function. Knowing that the free-space propagator satisfies the differential Eq.~(\ref{eqn:FreePhotonDiffA}) we apply the same differential operator to Eq.~(\ref{eqn:IntegralEq}), and after a short calculation we obtain the differential equation satisfied by the photon propagator in the half-space geometry,
\begin{eqnarray}
\left(\nabla_i\nabla_j-\delta_{ij}\nabla^2\right)\left[1+\theta(-z)\frac{K(\w)}{\epsilon_0}\right]D_{jk}(\br,\br';\w)\hspace{1.5 cm}\label{eqn:PropDiffEq}\\
-\w^2 D_{ik}(\br,\br';\w)=\frac{\epsilon_0}{(2\pi)^3}\int\rd^3 \bq\left(q_iq_k-\delta_{ik}\bq^2\right)e^{i\bq\cdot(\br-\br')}.\nonumber
\end{eqnarray}
The RHS can be re-written as
\begin{eqnarray}
-\frac{1}{(2\pi)^3}\int\rd^3 \bq\left(\bq^2\delta_{ik}-q_iq_k\right)e^{i\bq\cdot(\br-\br')}=
\nabla^2\delta^{\perp}_{ik}(\br-\br')\nonumber
\end{eqnarray}
where $\delta^{\perp}_{ik}(\br-\br')$ is the transverse delta-function. Now it is more apparent that the RHS of Eq.~(\ref{eqn:PropDiffEq}) is a distribution which, unlike the transverse delta function, is sharply localized around the point $\br=\br'$, because the non-local part of the transverse delta function is removed by the application of the Laplacian   as is obvious from the relation
\begin{equation}
-\nabla^2\left(\frac{1}{4\pi|\br-\br'|}\right)=\delta^{(3)}(\br-\br').
\end{equation}
The locality of $\nabla^2\delta^{\perp}_{ik}(\br-\br')$ is very helpful towards the solution of the differential Eq.~(\ref{eqn:PropDiffEq}), which is essentially a scattering problem. Its RHS contains a distribution representing a point-like source and our task is to work out reflection and transmission at the boundary of the dielectric. In order to proceed any further, we need to specify physical situation, i.e.~decide on which side of the boundary the source is placed. Since our ultimate aim is to work out the energy-shift in an atom located outside the dielectric, we choose to consider the case $z'>0$. Then we write Eq.~(\ref{eqn:PropDiffEq}) in a piecewise manner; on the vacuum side we have
\begin{eqnarray}
\left(\nabla_i\nabla_j-\delta_{ij}\nabla^2\right)D_{jk}(\br,\br';\w)\hspace{3.5 cm}\nonumber\\
-\w^2 D_{ik}(\br,\br';\w)=\epsilon_0\nabla^2\delta^{\perp}_{ik}(\br-\br'),\;\;\;z>0,\hspace{1 cm}\label{eqn:HalfspacePiecewise1}
\end{eqnarray}
and on the dielectric side we have
\begin{eqnarray}
\left(\nabla_i\nabla_j-\delta_{ij}\nabla^2\right)D_{jk}(\br,\br';\w)\hspace{3.5 cm}\nonumber\\
-\xi(\w)\w^2 D_{ik}(\br,\br';\w)=0,\;\;\;z<0,\hspace{1 cm}
\label{eqn:HalfspacePiecewise2}
\end{eqnarray}
with the behaviour of the propagator $D_{jk}(\br,\br';\w)$ across the interface $z=0$ is still to be determined. The local character of the RHS of Eq.~(\ref{eqn:PropDiffEq}) simplifies its solution in that it makes the RHS of Eq.~(\ref{eqn:HalfspacePiecewise2}) go to zero. 

In order to solve Eqs.~(\ref{eqn:HalfspacePiecewise1}) and (\ref{eqn:HalfspacePiecewise2}) we start with the following ansatz
\begin{eqnarray}
D_{ik}(\br,\br';\w)=
\left\{
\begin{array}{lr}
D_{ik}^{(t)}(\br,\br';\w) & z<0, \\
\; & \; \\
D_{ik}^{(0)}(\br-\br';\w)+D_{ik}^{(r)}(\br,\br';\w) & z>0.
\end{array}
\right. 
\nonumber\\\label{eqn:HSPropAnsatz1}
\end{eqnarray}
On the vacuum side we write the solution as a sum that consists of a particular solution $D_{ik}^{(0)}(\br-\br';\w)$, which we already know from Section \ref{sec:FreeSpacePhotProp}, Eq.~(\ref{eqn:PhotonFreeProp}), and a solution $D_{ik}^{(r)}(\br,\br';\w)$ of the corresponding homogeneous equation (i.e. Eq.~(\ref{eqn:HalfspacePiecewise1}) with the RHS set to zero) which represents the correction due to reflection at the boundary. The solution on the dielectric side $D_{ik}^{(t)}(\br,\br';\w)$ represents the transmitted part  and satisfies the homogeneous Eq.~(\ref{eqn:HalfspacePiecewise2}). The homogeneous solutions $D_{ik}^{(r)}(\br,\br';\w)$ and $D_{ik}^{(t)}(\br,\br';\w)$ are chosen in such a way that the general solution in Eq.~(\ref{eqn:HSPropAnsatz1}) satisfies appropriate electromagnetic boundary conditions across the interface $z=0$. To see what these boundary conditions should be recall the formal definition of the dressed propagator
\begin{equation}
D_{ij}(\br,\br';\w)=-\frac{i}{\hbar}\langle\Omega|T\left[D_i(\br,t)D_j(\br',t')\right]|\Omega\rangle.\label{eqn:AnotherPropDef}
\end{equation}
The displacement operator $D_i(\br,t)$ satisfies Maxwell's equations which follow from the Heisenberg equations of motion for the field operators. Therefore, the photon propagator, by virtue of its definition (\ref{eqn:AnotherPropDef}), when taken as a function of argument $\br$ and index $i$, is required to satisfy Maxwell's boundary conditions across the interface:
\begin{eqnarray}
\mathbf{E}_\parallel \mbox{\ continuous}\;&\longrightarrow &\ \epsilon^{-1}D_{\parallel j}\bigg|_{z=0^-}=D_{\parallel j}\bigg|_{z=0^+}\;\;\nonumber\\
D_z\mbox{\ continuous}\;&\longrightarrow &\ D_{zj}\bigg|_{z=0^-}=D_{z j}\bigg|_{z=0^+}\nonumber\\
\mathbf{B}_{\parallel}\mbox{\ continuous}\;&\longrightarrow &\ \epsilon^{-1}\nabla_z D_{\parallel j}\bigg|_{z=0^-}=\nabla_z D_{\parallel j}\bigg|_{z=0^+}\label{eqn:BoundaryConditions}
\end{eqnarray}
with $\parallel=\{x,y\}$. 

The apparent complication arising from the appearance of a non-standard distribution in the boundary-value problem (\ref{eqn:PropDiffEq}) is just an illusion. In fact, it is easier to find the solution of Eq.~(\ref{eqn:PropDiffEq}) than it is to solve the differential equation satisfied by the Green's function of the standard wave equation (see e.g.~\cite{Mills}). Eqs.~(\ref{eqn:HalfspacePiecewise1}) and (\ref{eqn:HalfspacePiecewise2}), together with the boundary conditions (\ref{eqn:BoundaryConditions}), form a boundary-value problem which is equivalent to the integral equation (\ref{eqn:IntegralEq}) with the choices $g(\br_1)=\theta(-z_1)$ (dielectric occupying the left half-space) and $z'>0$ (source located in vacuum).

In the following we shall use Eq.~(\ref{eqn:ZIntegratedProp1}) in the process of matching the boundary conditions. This is safe because we consider $z=0^\pm$ and the source located at $z'$ is always well away from the boundary, so that $z\neq z'$ is assured.

To proceed further we note that taking the divergence of the integral equation in Eq.~(\ref{eqn:IntegralEq}) and using the fact that the free-space propagator is transverse, $\nabla_iD^{(0)}_{ik}(\br-\br';\w)=0$, one infers that the dressed photon propagator is transverse everywhere as well
\begin{equation}
\nabla_iD_{ik}(\br,\br';\w)=0\label{eqn:DPropTransverse}.
\end{equation}
With this Eqs.~(\ref{eqn:HalfspacePiecewise1}) and (\ref{eqn:HalfspacePiecewise2}) simplify further  and partially Fourier transformed into the $(\bqp,z)$ space, cf. Eq.~(\ref{eqn:QZSpace}), may be written as
\begin{eqnarray}
\left(\nabla_z^2-\bqp^2+\w^2\right) D_{ij}(z,z')\hspace{3.6 cm}\nonumber\\
=\epsilon_0\left(\bqp^2-\nabla_z^2\right) \delta^\perp_{ij}(\bqp,z-z'),\;\;\;z>0 \hspace{1 cm}\\ \label{eqn:PropHelm1}
\left(\nabla_z^2-\bqp^2+\xi(\w)\w^2\right) D_{ij}(z,z')=0,\;\;\;z<0\hspace{1 cm}\label{eqn:PropHelm2}
\end{eqnarray}
where $\delta^\perp_{ij}(\bqp,z-z')$ is the Fourier transform of $\delta^\perp_{ij}(\br-\br')$ with respect to $\brp-\brp'$
\begin{equation}
\delta^\perp_{ij}(\bqp,z-z')=\int\rd^2(\brp-\brp')\: e^{-i\bqp\cdot(\brp-\brp')}\delta^\perp_{ij}(\br-\br').\nonumber
\end{equation}
Therefore the homogeneous solutions  $D_{ik}^{(r)}(\br,\br';\w)$ and $D_{ik}^{(t)}(\br,\br';\w)$ in Eq.~(\ref{eqn:HSPropAnsatz1}) must necessarily take the form
\begin{eqnarray}
D_{ij}^{(r)}(z,z')&=&-\frac{i\epsilon_0}{2}\left[R_{ij}(z')e^{ik_z z}+S_{ij}(z')e^{-ik_z z}\right],\;\;z>0,\nonumber\\\label{eqn:HomoGen1}\\
D_{ij}^{(t)}(z,z')&=&-\frac{i\epsilon_0}{2}\left[T_{ij}(z')e^{-ik_{zd} z}+U_{ij}(z')e^{ik_{zd}z}\right],\;\;z<0,\nonumber\\\label{eqn:HomoGen2}
\end{eqnarray}
with $k_z=\sqrt{\w^2-\bqp^2+i\eta}$ and $k_{zd}=\sqrt{\xi(\w)\w^2-\bqp^2}$ and the square roots taken such that $\mbox{Im}(k_z)\geq 0$ and $\mbox{Im}(k_{zd})\geq 0$. 
With this choice of sign for the square roots, the terms in Eqs.~(\ref{eqn:HomoGen1}) and (\ref{eqn:HomoGen2}) that contain exponentials $e^{-ik_z z}$ and $e^{ik_{zd} z}$ are unphysical as they represent waves that diverge at infinity. Thus we must set $S_{ij}=0=U_{ij}$. The remaining two matrices $R_{ij}$ and $T_{ij}$ are determined by the requirement that Eq.~(\ref{eqn:HSPropAnsatz1}) satisfies the boundary conditions in Eq.~(\ref{eqn:BoundaryConditions}). We note that, in addition, the transversality of the dressed propagator, Eq.~(\ref{eqn:DPropTransverse}), imposes rather stringent constraints on both $R_{ij}$ and $T_{ij}$. For example, the matrix $R_{ij}$ needs to be of the form
\begin{equation}
R_{ij}=v_i(\bqp) r_j(\bqp,z'),\label{eqn:DecompositionStart}
\end{equation}
where the vector $\mathbf{v}$ is such that $\mathbf{q}\cdot\mathbf{v}=0 $, leading to
\begin{equation}
\mathbf{v}=\left(v_x,v_y,-\dfrac{q_xv_y+q_yv_x}{k_z}\right),
\end{equation}
with $\mathbf{q}\equiv (\bqp, k_z)$. One might pick $v_x=-q_y$ and $v_y=q_x$ so that
\begin{equation}
\mathbf{v}=\left(-q_y,q_x,0\right).
\end{equation}
However, this choice is too restrictive on its own, as there is no \emph{a priori} reason for $D_{zj}$ to vanish. Therefore, an additional basis vector is needed in order to span the amplitude $R_{ij}$ in full generality. An obvious and convenient choice is to choose a vector that is orthogonal to both $\mathbf{q}$ and $\mathbf{v}$,
\begin{equation}
\mathbf{w}=\mathbf{v}\times\mathbf{q}=(q_xk_z,q_yk_z,-\bqp^2).
\end{equation}
Then we can represent $R_{ij}$ as the linear combination
\begin{eqnarray}
R_{ij}&=&\left[\alpha \mathbf{v}+\beta\mathbf{w}\right]_ir_j(z')\nonumber\\
&\equiv & e_i^{\TE}(k_z)r_j^{\TE}(z')+e_i^{\TM}(k_z)r_j^{\TM}(z').\nonumber
\end{eqnarray}
where we have recognized, apart from normalization factors, the transverse electric and transverse magnetic polarization vectors of Eq.~(\ref{eqn:PolarizationVec}). Similarly we have
\begin{equation}
T_{ij}= e_i^{\TE}(-k_{zd})t_j^{\TE}(z')+e_i^{\TM}(-k_{zd})t_j^{\TM}(z').\label{eqn:DecompositionEnd}
\end{equation}
We have chosen write out the $k_z$-dependence of the polarization vectors, even though $k_z$ and $k_{zd}$ are expressible in terms of the frequency $\w$ and the parallel wave-vector $\bqp$, because this explicitly indicates the wave-vector to which a given polarization vector is orthogonal to. The decomposition into transverse electric and transverse magnetic components significantly simplifies the matching of boundary conditions. The dressed photon propagator can now be written in the form
\begin{eqnarray}
D_{ij}(z,z')=-\frac{i\epsilon_0}{2}\sum_{\lambda}
\bigg\{
\left[
e^\lambda_i(-k_{zd})t_j^\lambda e^{-ik_{zd}z}
\right]\theta(-z)\hspace{1.2 cm}\nonumber\\
+\bigg[
e^\lambda_i(k_{z})r_j^\lambda e^{ik_{z}z}+\frac{\w^2}{k_z}e^\lambda_i(-k_{z})e^\lambda_j(-k_{z})e^{-ik_{z}(z-z')}
\bigg]\theta(z) 
\bigg\},\nonumber\\
\label{eqn:PropBoundCondMatch}
\end{eqnarray}
the last term of which is the free-space photon propagator from Eq. (\ref{eqn:ZIntegratedProp1}) for $z-z'<0$, as appropriate for the matching of boundary conditions at $z=0$ when $z'>0$. Imposing the boundary conditions of Eq.~(\ref{eqn:BoundaryConditions}) we find that
\begin{eqnarray}
r_j^\lambda &=& r_\lambda\;e_j^\lambda(-k_z)\frac{\w^2}{k_z}e^{ik_z z'},\nonumber\\
t_j^\lambda &=& t_\lambda\; e_j^\lambda(-k_z)\frac{\xi(\w)\w^2}{k_z}e^{ik_z z'},\nonumber
\end{eqnarray}
with $r_\lambda$ and $t_\lambda$ being the standard Fresnel's reflection and transmission coefficients listed in Eq.~(\ref{eqn:Fresnel}). Plugging the above amplitudes into Eq.~(\ref{eqn:PropBoundCondMatch}) we then readily obtain the photon propagator given in Eq.~(\ref{eqn:HSPropFinal}).

We can readily apply the same methods to obtain the propagator in the case when the source is placed in the dielectric, i.e.~for $z'<0$. The calculation goes along exactly the same lines as for $z'>0$ and one can show that the photon propagator in the case of the source being placed in the dielectric is given by
\begin{widetext}
\begin{eqnarray}
D_{ij}(\br,\br';\w)=\theta(-z)D^{(\epsilon)}_{ij}(\br-\br';\w)-\frac{i\epsilon_0}{(2\pi)^2}\sum_{\lambda}\int\rd_2\bqp \frac{\xi(\w)\w^2}{2k_{zd}} e^{i\bqp\cdot(\brp-\brp')}\hspace{5 cm}\nonumber\\
\times\bigg\{
\theta(-z)\left[\xi(\w)e^\lambda_i(\bqp,-k_{zd})e_j^\lambda(\bqp,k_{zd})r_L^\lambda\right]e^{-ik_{zd}(z+z')}+\theta(z)\left[e^\lambda_i(\bqp,k_{z})e_j^\lambda(\bqp,k_{zd})t_L^\lambda \right]e^{ik_{z}z-ik_{zd}z'}\bigg\},\;\;\;\;\;\;.\label{eqn:HSPropFinal2}
\end{eqnarray}
\end{widetext}
Here the reflection and transmission coefficients are those appropriate for left-incident modes; they are given by
\begin{eqnarray}
r_L^\TE=\frac{k_{zd}-k_z}{k_z+k_{zd}},\;\;\;r_L^\TM=\frac{k_{zd}-\xi(\w)k_z}{\xi(\w)k_z+k_{zd}},\nonumber\\
t_L^\TE=\frac{2k_{zd}}{k_z+k_{zd}},\;\;\;t_L^\TM=\frac{2\sqrt{\xi(\w)}k_{zd}}{\xi(\w)k_z+k_{zd}}.
\label{eqn:FresnelA}
\end{eqnarray}
It is easily verified that $D_{ij}(\br,\br';\w)$ is indeed transverse everywhere.

\section{Simple model for $\epsilon(\w)$}\label{App:Epsilon}
In order to determine the dielectric permittivity of our model we use the equations of motion for the fields that follow from the Hamiltonian (\ref{eqn:EMHam})--(\ref{eqn:PolEMCoupling}) and the commutation relations (\ref{eqn:EMCommutator})--(\ref{eqn:BathCommutator}):
\begin{eqnarray}
\frac{\partial }{\partial t}\mathbf{D}(\br,t)&=&\frac{1}{\mu_0}\boldsymbol{\nabla}\times \mathbf{B}(\br,t)\label{eqn:EqnOfMtn1}\\
\frac{\partial  }{\partial t}\mathbf{B}(\br,t)&=&-\boldsymbol{\nabla}\times \mathbf{E}(\br,t)\label{eqn:EqnOfMtn2}\\
\frac{\partial}{\partial t}\mathbf{X}(\br,t)&=&\frac{1}{\mathcal{M}}\mathbf{P}(\br,t)\label{eqn:EqnOfMtn3}\\
\frac{\partial}{\partial t}\mathbf{P}(\br,t)&=&-\mathcal{M}\w_{\rm T}^2\mathbf{X}(\br,t) + g(\br) \mathbf{E}(\br,t)\nonumber\\ && +\int_0^\infty\rho_\nu\nu^2\mathbf{Y}_\nu(\br,t)\label{eqn:EqnOfMtn4}\\
\frac{\partial}{\partial t}\mathbf{Y}_\nu(\br,t)&=&\frac{1}{\rho_{\nu}}\mathbf{Z}_{\nu}(\br,t)\label{eqn:EqnOfMtn5}\\
\frac{\partial}{\partial t}\mathbf{Z}_{\nu}(\br,t)&=&-\rho_\nu \nu^2 \mathbf{Y}_\nu (\br,t) + \rho_\nu \nu^2\mathbf{X}(\br,t)\label{eqn:EqnOfMtn6}
\end{eqnarray}
First we deal with the subsystem consisting of the polarization field and the reservoir. It is well known \cite{Langevin} that when a quantized harmonic oscillator is coupled to a bath its equation of motion takes the form of a quantum Langevin equation. Thus we expect the equation of motion for the polarization field, which is nothing but a set of independent oscillators, to take the form
\begin{eqnarray}
\mathcal{M}\frac{\partial ^2}{\partial t^2}\mathbf{X}(\br,t)&+&\int_{-\infty}^t\rd t' \mu(t-t')\frac{\partial}{\partial t'}\mathbf{X}(\br, t')\nonumber\\
&+&\mathcal{M}\w_{\rm T}^2\mathbf{X}(\br, t)=\mathbf{F}_{\rm ran}(\br, t)+\mathbf{F}_{\rm ext}(\br,t)\;\;\nonumber\\
\label{eqn:QuantumLangevin}
\end{eqnarray}
where  $\mu(t-t')$ is the so-called memory function related to dissipation and $\mathbf{F}_{\rm ran}(\br,t)$ represents some random force operator (see e.g.~\cite{Langevin} for details). Both $\mu(t-t')$ and $\mathbf{F}_{\rm ran}(\br,t)$ arise as a consequence of the coupling to the bath and are to be determined in terms of the parameters of our model. The term $\mathbf{F}_{\rm ext}(t)$ represents any external forces (i.e.~those in addition to the harmonic restoring force) that may be applied to the polarization field. To show that Eqs.~(\ref{eqn:EqnOfMtn3})--(\ref{eqn:EqnOfMtn6}) indeed combine to yield an equation of the form of Eq.~(\ref{eqn:QuantumLangevin}), we eliminate $\mathbf{P}(\br,t) $ and $\mathbf{Z}_{\nu}(\br,t) $ and rewrite the equations for $\mathbf{X}(\br,t) $ and $\mathbf{Y}_\nu(\br,t) $ as 
\begin{eqnarray}
\bigg(\frac{\partial^2}{\partial t^2} + \w_{\rm T}^2 + \frac{1}{\mathcal{M}}\int_0^\infty \rd \nu &&\hspace*{-4mm} \rho_\nu  \nu^2
 \bigg) \mathbf{X}(\br,t)\nonumber\\
&=&\frac{1}{\mathcal{M}}\int_0^\infty \rd \nu \rho_\nu \nu^2\mathbf{Y}_\nu(\br,t),\;\;\;\label{eqn:EqnOfMtn3plus4}\\
\left(\frac{\partial^2}{\partial t^2} + \nu^2 \right)\mathbf{Y}_\nu(\br,t) &=& \nu^2\mathbf{X}(\br,t)\label{eqn:EqnOfMtn5plus6}.
\end{eqnarray}
The most general solution of Eq.~(\ref{eqn:EqnOfMtn5plus6}) may be written as
\begin{eqnarray}
\mathbf{Y}_{\nu}(\br, t)=\mathbf{Y}^{H}_{\nu}(\br, t)+\int_{-\infty}^{\infty}\rd t' G_\nu(t-t')\mathbf{X}(\br,t)\;\;\;\;\label{eqn:GenSolution}
\end{eqnarray}
where $\mathbf{Y}^{H}_{\nu}(\br, t)$ is the solution of the homogeneous equation, i.e.~Eq.~(\ref{eqn:EqnOfMtn5plus6}) with its RHS set to zero,
\begin{eqnarray}
\mathbf{Y}^{H}_{\nu}(\br, t)=\mathbf{Y}_{\nu}(\br,0)\cos(\nu t) +\frac{\mathbf{Z}_{\nu}(\br,0)}{\rho_\nu}\frac{\sin(\nu t)}{\nu}.\label{eqn:HomoSol}\;\;\;
\end{eqnarray}
We assume that bath operators $\mathbf{Y}_{\nu}(\br,0)$ and $\mathbf{Z}_{\nu}(\br,0)$ satisfy the canonical commutation relations at the initial time $t=0$, cf.~Eq.~(\ref{eqn:BathCommutator}), which we take as a moment in the distant past when the interaction has been switched on. The second term in Eq.~(\ref{eqn:GenSolution}) is a particular solution expressed in terms of the Green's function
\begin{equation}
G_\nu(t-t')=\frac{1}{2\pi}\int_{-\infty}^\infty\rd \w \frac{\nu^2}{\nu^2-(\w+i\epsilon)^2}e^{-i\w(t-t')}\label{eqn:OscillatorGreens}
\end{equation} 
The $i\epsilon $ prescription for handling the pole ensures that we have a retarded Green's function with $G_\nu(t-t')=0$ for $t-t'<0$. Eq.~(\ref{eqn:OscillatorGreens}) is easily obtained from Eq.~(\ref{eqn:EqnOfMtn5plus6}) by using Fourier transforms according to
\begin{equation}
\mathbf{Y}_{\nu}(\br,\w)=\int_{-\infty}^{\infty}\rd t e^{i\w t}\mathbf{Y}_{\nu}(\br,t).
\end{equation}
Note that the choice of the retarded solution breaks time reversal invariance, as has been noted in Ref.~\cite{Langevin}. The integral in Eq.~(\ref{eqn:OscillatorGreens}) is easily worked out using the residue theorem, and Eq.~(\ref{eqn:GenSolution}) may be rewritten as
\begin{eqnarray}
\mathbf{Y}_{\nu}(\br, t) &=& \mathbf{Y}^{H}_{\nu}(\br, t)+\mathbf{X}(\br,t)\nonumber\\
&-&\int_{-\infty}^{t}\rd t' \cos[\nu(t-t')]\frac{\partial}{\partial t}\mathbf{X}(\br,t)\;\;\label{eqn:GenSolution2}
\end{eqnarray}
where we have integrated by parts. Plugging the above expression into Eq.~(\ref{eqn:EqnOfMtn3plus4}) we obtain
\begin{eqnarray}
\mathcal{M}\frac{\partial ^2}{\partial t^2}\mathbf{X}(\br,t)\hspace{5.5 cm}\nonumber\\
+\int_{-\infty}^t\rd t' \left\{\int_0^\infty\rd \nu\rho_\nu\nu^2\cos[\nu(t-t')] \right\}\frac{\partial}{\partial t'}\mathbf{X}(\br, t')\nonumber\\
+\mathcal{M}\w_{\rm T}^2\mathbf{X}(\br, t)=\int_0^\infty \rd \nu \rho_\nu \nu^2 \mathbf{Y}_{\nu}^H(\br,t).\hspace{0.5 cm}\label{eqn:QuantumLangevinOurModel}
\end{eqnarray}
This is the quantum Langevin equation that follows from our model. Comparing with Eq.~(\ref{eqn:QuantumLangevin}) lets us to identify 
\begin{eqnarray}
\mathbf{F}_{\rm ran}(\br, t)&=&\int_0^\infty\rd \nu \rho_\nu \nu^2 \mathbf{Y}^H_\nu(\br,t),\label{eqn:RandomForce}\\
\mu(t-t')&=&\int_0^\infty\rd\nu\rho_\nu\nu^2\cos[\nu(t-t')].
\end{eqnarray}
Now we are in the position to choose the bath oscillator masses $\rho_{\nu}$.
Having in mind a simple single-resonance model of the dielectric permittivity, we choose $\rho_{\nu}$ in such a way that the friction term in Eq.~(\ref{eqn:QuantumLangevinOurModel}) is local in time i.e.~it is non-vanishing only for $t=t'$. This is achieved by choosing
\begin{equation}
 \rho_\nu=\frac{4\mathcal{M}\gamma}{\pi\nu^2}\label{eqn:rho_nu}
\end{equation}
which gives a frequency-independent coupling between the bath and polarization oscillators, cf.~Eq.~(\ref{eqn:BathResCoupling}). Then Eq.~(\ref{eqn:QuantumLangevinOurModel}) becomes
\begin{eqnarray}
\frac{\partial ^2}{\partial t^2}\mathbf{X}(\br,t)&+&2\gamma\frac{\partial}{\partial t}\mathbf{X}(\br, t)+\w_{\rm T}^2\mathbf{X}(\br, t)\nonumber\\
&=&\frac{1}{\mathcal{M}}\int_0^\infty \rd \nu \rho_\nu \nu^2 \mathbf{Y}_{\nu}^H(\br,t)+\frac{g(\br)}{\mathcal{M}}\mathbf{E}(\br, t).\hspace{0.9 cm}\label{eqn:QuantumLangevinOurModel2}
\end{eqnarray}
We have augmented this equation by the ``external force'' term that arises when the polarization field is coupled to the electromagnetic field, which according to Eqs.~(\ref{eqn:EqnOfMtn1})--(\ref{eqn:EqnOfMtn2}) satisfies the equation of motion
\begin{equation}
\boldsymbol{\nabla}\times\left[\boldsymbol{\nabla}\times\mathbf{E}(\br,t)\right]+\mu_0\epsilon_0\frac{\partial^2}{\partial t^2}\mathbf{E}(\br,t)=
-\mu_0 g(\br) \frac{\partial^2}{\partial t^2}\mathbf{X}(\br,t)\label{eqn:WaveForE}
\end{equation}
with $\mathbf{D}(\br,t)=\epsilon_0\mathbf{E}(\br,t)+g(\br)\mathbf{X}(\br, t)$.  Similarly to the reservoir field discussed before, the most general solution of Eq.~(\ref{eqn:QuantumLangevinOurModel2}) is given as a sum of the homogeneous solution (i.e.~the solution of Eq.~(\ref{eqn:QuantumLangevinOurModel2}) with the RHS set to zero and the assumption that the oscillators are underdamped) and the particular solution. The homogeneous solution is of the same form as Eq.~(\ref{eqn:HomoSol}) except for an additional damping factor proportional to $e^{-\gamma t}$. Since we assume that the initial time is a moment in the distant past we may discard the homogeneous solution which is exponentially small for $\gamma t \gg 1$. The particular solution is easily obtained in Fourier space and is given by
\begin{equation}
\mathbf{X}(\br, t)=\frac{1}{2\pi\mathcal{M}}\int_{-\infty}^\infty\rd \w\frac{\mathbf{F}_{\rm ran}(\br,\w)+g(\br)\mathbf{E}(\br,\w)}{\w_{\rm T}^2-\w^2+2i\gamma\w}e^{-i\w t}\;\;\;\label{eqn:ParticularSol}
\end{equation}
where $\mathbf{F}_{\rm ran}(\br,\w)$ is the Fourier transform of Eq.~(\ref{eqn:RandomForce}) and is given explicitly by
\begin{equation}
\mathbf{F}_{\rm ran}(\br,|\w|)=4\gamma\mathbf{Y}_{|\w|}(\br,0)+i\frac{\pi}{\mathcal{M}}|\w|\mathbf{Z}_{|\w|}(\br,0)\label{eqn:RandomForce2}.
\end{equation}
Substitution of the solution (\ref{eqn:ParticularSol}) into Eq.~(\ref{eqn:WaveForE}) yields
\begin{eqnarray}
\boldsymbol{\nabla}&\times&\left[\boldsymbol{\nabla}\times\mathbf{E}(\br,\w)\right]\nonumber\\
&-&\mu_0\epsilon_0\w^2\left[1+\frac{g^2(\br)}{\mathcal{M}\epsilon_0}\frac{1}{\w_{\rm T}^2-\w^2-2i\gamma\w}\right]\mathbf{E}(\br,\w)\nonumber\\
&=&\mu_0\frac{g(\br)}{\mathcal{M}}\frac{\mathbf{F}_{\rm ran}(\br,\w)}{\w_{\rm T}^2-\w^2-2i\gamma\w}.
\end{eqnarray}
We may now read off the dielectric function given by
\begin{equation}
 \frac{\epsilon}{\epsilon_0}=1+g^2(\br)\frac{\w^2_{\rm P}}{\w_{\rm T}^2-\w^2-2i\gamma\w}.
\end{equation}
with $\w_{\rm P}^2\equiv 1/\mathcal{M}\epsilon_0$. The quantity that appears on the RHS is proportional to the so called noise-current operator which is introduced \emph{ad hoc} in the phenomenological formulation of the quantum theory developed in Ref.~\cite{PhenQED}. In fact we have
\begin{equation}
\mathbf{J}_{\rm N}(\br, \w)= -i\w \frac{g(\br)}{\mathcal{M}}\frac{\mathbf{F}_{\rm ran}(\br,\w)}{\w_{\rm T}^2-\w^2-2i\gamma\w}.
\end{equation}
Since the operator $\mathbf{F}_{\rm ran}(\br,\w)$ depends only on the initial coordinates and momenta of the bath, cf. Eq. (\ref{eqn:RandomForce2}), for which the commutation relations are known, it is relatively easy to verify that
\begin{eqnarray}
\left[J_i(\br,\w),J^\dagger_k(\br',\w')\right]=4\pi\hbar\epsilon_0\mbox{Im}[\epsilon(\br,\w)]\w^2\delta^{(3)}(\br-\br')\nonumber\\
\times\delta(\w-\w')\delta_{ik}\;.\hspace*{10mm}
\end{eqnarray}
This derivation justifies these phenomenologically introduced commutation rules on a microscopic level.
We just note that this result differs from Eq.~(\ref{eqn:YenaCommutator}) to be used in the following Section by a factor of $(2\pi)^2$ due to a different definition of the Fourier transform.

\section{Photon propagator from phenomenological QED}\label{App:DressPropPhen}
The phenomenological theory of quantum electrodynamics, as developed in Ref.~\cite{PhenQED}, gives the electric field operator as
\begin{equation}
 E_i(\br,t)=-i\mu_0\int\rd^3\br'\int_0^\infty\rd\w e^{-i\w t}G_{ik}(\br,\br';\w)J_k(\br',\w)+{\rm H.C.}\label{eqn:YenaEFieldOP}
\end{equation}
where $J_k(\br,\w)$ is the so-called noise current operator satisfying the following commutation relation
\begin{eqnarray}
 \left[J_i(\br,\w),J^\dagger_k(\br',\w')\right]=\frac{\hbar\epsilon_0}{\pi}\mbox{Im}[\epsilon(\br,\w)]\w^2\delta^{(3)}(\br-\br')\nonumber\\
 \times\delta(\w-\w')\delta_{ik} \;,\hspace*{5mm}\label{eqn:YenaCommutator}
\end{eqnarray}
and $G_{ik}(\br,\br';\w)$ is the Green's function of the wave equation satisfying
 \begin{eqnarray}
(\nabla_i\nabla_j-\delta_{ij}\nabla^2)G_{jk}(\br,\br';\w)-\epsilon(\br,\w)\w^2G_{ik}(\br,\br';\w)
\nonumber\\=\delta_{ik}\delta^{(3)}(\br-\br'),\label{eqn:GreensTensor}\hspace{1 cm}
 \end{eqnarray}
with the additional requirement that it is retarded in time. Note, however, that there is no transversality condition imposed and the RHS of Eq.~(\ref{eqn:GreensTensor}) is just a diagonal $\delta$ function. For an overview of the noise-current approach and some applications see Ref.~\cite{Scheel}. In the following we shall use two properties of the Green's tensor in particular, its reciprocity
\begin{equation}
G_{ik}(\br,\br';\w)=G_{ki}(\br',\br;\w)\;,
\end{equation}
and the integral relation
\begin{eqnarray}
\int\rd^3\br''\;\w^2\:\mbox{Im}[\epsilon(\br,\w)] G_{ik}^*(\br,\br'';\w) G_{jk}(\br',\br'';\w) \nonumber\\
= \mbox{Im}[G_{ij}(\br,\br';\w)].\label{eqn:GreensIntegral}
\end{eqnarray}
To prove the latter one multiplies Eq.~(\ref{eqn:GreensTensor}) from the left by $G_{mi}^*(\br'',\br;\w)$ and integrates over $\br$. Then taking the difference between the resulting relation and its complex conjugate integrated by parts yields Eq.~(\ref{eqn:GreensIntegral}).

In order to calculate the Feynman propagator of the electric field operator, i.e.~the quantity
\begin{equation}
 D^{\rm E}_{ij}(\br,\br',t,t')=-\frac{i}{\hbar}\langle 0|T\left[E_i(\br,t)E_j(\br',t')\right]|0\rangle,
\end{equation}
we substitute into the above definition the operator (\ref{eqn:YenaEFieldOP}) and use Eqs.~(\ref{eqn:YenaCommutator})--(\ref{eqn:GreensIntegral}). We arrive at
\begin{eqnarray}
D^{\rm E}_{ij}(\br,\br',t,t')=-\frac{i}{\pi\epsilon_0}\int_0^\infty\rd\w\:\w^2 \left[\theta(t-t')e^{-i\w(t-t')}\right.
\nonumber\\
\left. +\theta(t'-t)e^{i\w(t-t')}\right]\mbox{Im}[G_{ij}(\br,\br';\w)].\nonumber
\end{eqnarray}
Now we carry out the Fourier transform with respect to $t-t'$ using the distributional identities
\begin{equation}
 \int_0^\infty\rd\tau e^{\pm i \tau \Omega}=\pi \delta(\Omega)\pm i \frac{\mathcal{P}}{\Omega}\;,
\end{equation}
where $\mathcal{P}$ denotes the Cauchy principal value, and obtain
\begin{eqnarray}
 D^{\rm E}_{ij}(\br,\br';\Omega)=\frac{2}{\pi\epsilon_0}\mathcal{P}\int_0^\infty\rd\w\frac{\w^3}{\Omega^2-\w^2}\mbox{Im}[G_{ij}(\br,\br';\w)]\hspace{0.7 cm}\nonumber\\
-\frac{i}{\epsilon_0}\int_0^\infty\rd\w\w^2\left[\delta(\Omega-\w)+\delta(\Omega+\w)\right]\mbox{Im}[G_{ij}(\br,\br';\w)].\hspace{.5 cm}\label{eqn:YenaPropA}
\end{eqnarray}
The Green's tensor must satisfy retarded boundary conditions in time in order to preserve causality. This means that it is analytical in the upper half of the complex $\w$ plane. Analyticity in the upper-half of the $\w$-plane leads to Kramers-Kronig relations \cite{Jackson}, so that the Green's tensor inherits the causality properties of the permittivity. In particular, its imaginary part is an odd function of frequency $\w$, whereas its real part is even in $\w$. With that we can proceed to deal with the principal-value integral in Eq.~(\ref{eqn:YenaPropA}). Since $\mbox{Im}[G_{ij}(\br,\br';-\w)]=-\mbox{Im}[G_{ij}(\br,\br';\w)]$ and the remaining part of the integrand is also odd, we extend the lower integration limit to $-\infty$ and compensate by multiplying by $1/2$. On the other hand, the real part of the Green's tensor is even in $\w$, so that we can replace
\begin{equation}
\mbox{Im}[G_{ij}(\br,\br',\w)]\rightarrow \frac{1}{i}G_{ij}(\br,\br',\w)
\end{equation}
without changing the value of the integral. Thus the principal-value integral in Eq.~(\ref{eqn:YenaPropA}) becomes
\begin{equation}
\frac{\mathcal{P}}{i\pi\epsilon_0}\int_{-\infty}^\infty\rd\w\frac{\w^3}{\Omega^2-\w^2}G_{ij}(\br,\br';\w).\label{eqn:PrincipalPart}
\end{equation}
To work out this integral we consider a contour of integration $\gamma$ that runs from $-\infty$ to $\infty$ and above the poles at $\w=\pm\Omega$ and then closes up in the upper half of the $\w$-plane along the large semicircle $|\w|\rightarrow\infty$. Because the Green's tensor in analytic in the upper half-plane, the such calculated integral vanishes and we can express the principal-value integral as
\begin{equation}
 \mathcal{P}\int=-\int_{\gamma^-}-\int_{\gamma^+}-\int_{\Gamma}
\end{equation}
where $\gamma^{\pm}$ denotes the clockwise contours that go around the poles at $\w=\pm\Omega$ respectively and $\Gamma$ denotes the contribution from the large semicircle taken counter-clockwise. Using the residue theorem we derive that the contribution from $\gamma^\pm$ is given by
\begin{equation}
 -\frac{1}{\epsilon_0}\Omega^2 G'_{ij}(\br,\br';\Omega),
\end{equation}
The large semicircle $\Gamma$ contributes the delta function
\begin{equation}
 -\frac{1}{\epsilon_0}\delta_{ij}\delta^{(3)}(\br-\br'),
\end{equation}
for whose calculation we have used the fact that asymptotically the Green's tensor behaves as \cite{Scheel}
\begin{equation}
 \lim_{|\w|\rightarrow\infty}\w^2 G_{ij}(\br,\br;\w)=-\delta_{ij}\delta^{(3)}(\br-\br').
\end{equation}
The $\delta$ function integral in Eq.~(\ref{eqn:YenaPropA}) is easily seen to be
\begin{equation}
 -\frac{i}{\epsilon_0}\Omega^2\;\mbox{Im}[G_{ij}(\br,\br';|\Omega|)],
\end{equation}
so that the final result for the relation between the photon propagator and the Green's function of the wave equation on the real $\Omega$-axis can be compactly written as
\begin{equation}
 D^{\rm E}_{ij}(\br,\br';\Omega)=-\frac{\Omega^2}{\epsilon_0}G_{ij}(\br,\br';|\Omega|)-\frac{1}{\epsilon_0}\delta_{ij}\delta^{(3)}(\br-\br').\label{eqn:YenaPropFinal}
\end{equation}
A similar formula has been given in Ref.~\cite{Tomas}. We would like to use this result for a comparison with the results of Section \ref{sec:PhotonDress}. First we need to emphasize that what we have calculated here is the propagator for the electric field $\mathbf{E}$, whereas Section \ref{sec:PhotonDress} derives the propagator for the displacement field $\mathbf{D}$. Therefore, the results can coincide only when $\br$ and $\br'$ are both located outside the dielectric, which is why we restrict ourselves to this case. Then the Green's tensor $G_{ij}(\br,\br';\w)$ splits into a free-space part $G^{(0)}_{ij}$ and a correction $G^{(r)}_{ij}$ that describes the reflection of the electromagnetic field from the surface, and Eq.~(\ref{eqn:YenaPropFinal}) can be rewritten as
\begin{eqnarray}
 D^{\rm E}_{ij}(\br,\br';\Omega)=-\frac{\Omega^2}{\epsilon_0}\left[G^{(0)}_{ij}(\br-\br';|\Omega|)+\delta_{ij}\delta^{(3)}(\br-\br')\right]\nonumber\\
 -\frac{\Omega^2}{\epsilon_0}G^{(r)}_{ij}(\br,\br';|\Omega|).\hspace{1 cm}\label{eqn:YenaPropFinal1}
\end{eqnarray}
This makes clear that the Feynman propagator is an even function of $\Omega$, unlike the Green's function of the wave equation which has the same analytical structure as the dielectric function.
It is not difficult to verify that for the particular geometry considered here, the dielectric half-space, Eq.~(\ref{eqn:YenaPropFinal1}) indeed holds. The terms in square brackets combine to deliver the \emph{transverse} free-space propagator as given in Eq.~(\ref{eqn:PhotonFreeProp}). The reflected part $ G^{(r)}_{ij}(\br,\br';|\Omega|)$, which can be found e.g.~in Ref.~\cite{PhenQED}, satisfies the homogeneous wave equation. Therefore, it is automatically transverse 
\begin{equation}
\nabla_i G^{(r)}_{ij}(\br-\br';\w)=0
\end{equation}
and for real $\w$ it coincides with the reflected part of the photon propagator $D^{\rm E}_{ij}(\br,\br';\w)$ given in Eq.~(\ref{eqn:HSPropFinal}), though away from the real axis they are different due to the different boundary conditions in time. $D^{\rm E}_{ij}(\br,\br';t-t')$ is a Feynman propagator whereas $G_{ij}(\br,\br',t-t')$ gives the retarded solutions of the wave equation.


\begin{thebibliography}{99}
\bibitem{fn1}{All the quantum fields here and in the following are of course time-dependent, though we suppress this time dependence in the arguments for brevity.}
\bibitem{fn2}{At this stage the analogy between the quantity $(\epsilon_0\mathcal{M})^{-1}$ and the plasma frequency, as commonly introduced in the free-electron model of a metal is not readily apparent. This choice of notation justifies itself once the precise form of $\rho_\nu$ in Eqs. (\ref{eqn:PolarizationHam2})-(\ref{eqn:FrequencyShift2}) is chosen so that the dielectric function of the model emerges (see Appendix \ref{App:Epsilon} for details). }
\end{thebibliography}
\end{document}